\documentstyle[multicol,epsf,aps,prb,floats]{revtex}
\begin{document}

\title{Two-Channel Kondo Physics from Tunnelling
  Impurities with Triangular Symmetry.}
\author{Aris L. Moustakas and Daniel S. Fisher}
\address{Department of Physics, Harvard University, Cambridge, MA 02138}
\maketitle

\begin{abstract}
Tunnelling impurities in metals have been known for some time to have
the potential for exhibiting Kondo-like physics. However previous
models based on an impurity hopping between two equivalent positions
have run into trouble due to the existence of relevant operators that
drive the system away from the non-Fermi-liquid Kondo fixed point. In
the case of an impurity hopping among positions with higher symmetry,
such as triangular symmetry, it is shown here that the
non-Fermi-liquid behavior at low temperatures can be generic.  Using
various bosonization techniques, the fixed point is shown to be {\em
  stable}. However, unlike the conventional two-channel Kondo (2CK)
model, it has {\em four} leading irrelevant operators, implying that
while the form of the singular temperature dependence of physical
quantities is similar to the 2CK model, there will not be simple
universal amplitude ratios.  The phase diagram of this system is
analyzed and a critical manifold is found to separate the
non-Fermi-liquid from a conventional Fermi liquid fixed point.
Generalization to higher symmetries, such as cubic, and the
possibility of physical realizations with dynamic Jahn-Teller
impurities is discussed.
\end{abstract}

\section{Introduction}
\label{sec:Ic3v}

Following pioneering work by Nozi\`{e}res and Blandin\cite{Nozieres1},
the multichannel Kondo model has attracted much attention because of
the prospect of novel, low temperature behavior. Various theoretical
methods have been developed to study the nature of the zero
temperature Kondo fixed points for a variety of cases, in particular
with an arbitrary number of
channels\cite{Andrei,Affleck1,Affleck3,Affleck1a,EK,Cragg} and with either
one or two\cite{Affleck5,Jones2,Senguptathanks,Jones3} Kondo impurities.
Despite various intriguing predictions, no experimental realization of
the multichannel Kondo effect has yet been conclusively demonstrated.
The difficulty lies in the fact that the non--Fermi--liquid fixed
points are unstable to various symmetry--breaking processes, such as
channel anisotropy, which turn out to be present in real experimental
situations. For example, Nozi\`{e}res and Blandin\cite{Nozieres1}
pointed out that anisotropy between the channels, caused by lattice
effects, would destroy the non--Fermi--liquid ground state.

Subsequently, Vlad\'{a}r and Zawadowski\cite{Zawadowski1} were the
first to suggest a different realization of the two-channel Kondo
model, with the role of the spins and the channels of the conventional
magnetic model interchanged. In this case, a {\em non-magnetic}
impurity tunnelling between two sites plays the role of the magnetic
spin-1/2 impurity in the original Kondo formulation with electrons
coupled orbitally to the impurity, while the spin index of the
electrons is a spectator, playing the role of the channel index.
Therefore, since in the absence of a magnetic field in the magnetic
Kondo system the spin-up and spin-down electrons are degenerate,
channel anisotropy is no longer a problem.  This led Vlad\'{a}r and
Zawadowski\cite{Zawadowski1} to predict that such a system should
exhibit non--Fermi--liquid behavior at low temperatures,
characteristic of the two-channel Kondo (2CK) model. Recently, low
temperature tunnelling data by Buhrman {\em et
  al}\cite{Ralph1,Ralph2,Buhrman}
in very small metallic contacts have been interpreted by Buhrman {\em et
  al}\cite{Ralph1,Ralph2,Buhrman} in terms of two-channel Kondo-like
physics. Their measurements are claimed to be consistent with certain
exact results for the 2CK model obtained by Affleck and
Ludwig,\cite{Affleck1,Affleck3,Affleck1a} but at this point, the
interpretation is controversial.\cite{Wingreen,MF2} Again, the
problem arises from the absence of a symmetry which prevents other
relevant terms from appearing; the most important of these is bare
impurity tunnelling, which in the conventional Kondo model corresponds
to a magnetic field on the impurity. In a previous paper\cite{MF2}, we
studied this system in detail and argued that it is very
unlikely that 2CK physics could be observable over a substantial
temperature range in such a two-site impurity system.

A way to circumvent the problem of the appearance of relevant
operators is to impose symmetry conditions on the orbital degrees of
freedom so that the relevant operators which plagued the
two-site impurity system are not allowed to appear by symmetry.  The
first question that naturally arises is what are the minimum symmetry
requirements at some high (or intermediate) energy scale so that the
system will flow under renormalization (RG) towards the non-Fermi
liquid fixed point, {\em without} any fine tuning. Clearly, the low
temperature non-Fermi liquid behavior one should be primarily aiming
for, is that of the two-channel Kondo  model due to its relative
simplicity.

For tunnelling non-magnetic impurities, the ``channel'' (i.e. spin)
symmetry is ensured, in the absence of a magnetic
field, due to time reversal invariance. As for the symmetry of the
impurity motion, an essential requirement is that at intermediate
energies the impurity, dressed with high energy electrons, should
behave like a degenerate doublet, just as the impurity
spin-$\frac{1}{2}$ does in the conventional 2CK model. For this to be
true without any parameter tuning, the symmetry group of the system has to
allow for  degenerate doublets to occur, i.e. it has to have a
two dimensional representation. The simplest group with such a
property is the triangular group $C_{3v}$, which, with six elements, is
the smallest non-abelian group.

In this paper we are thus led to consider a non-magnetic impurity in a
metal that can tunnel between the corners of an equilateral triangle.
In this case, the left and right moving impurity states are related by
symmetry and are thus degenerate. Not surprisingly, in order for the
system to be in the basin of attraction of the 2CK point, we must
require that at some intermediate energy, the left-right doublet of
the impurity is the lowest energy orbital state of the impurity.
Naively one might expect that this is impossible, since the ground
state of a particle in an external potential is very generally
non-degenerate. However, in a solid the impurity interacts with other
electrons, including localized atomic electrons, whose state is
affected by the impurity motion. By taking these interaction effects
into account, we explicitly demonstrate in some simple and potentially
physically relevant examples, that it is indeed possible for the
ground state of the impurity-plus-high-energy-electrons complex to be
degenerate.

We then analyze the weak electron-impurity interaction limit,
determine the criterion for flowing under the renormalization group
(RG) toward the 2CK fixed point, and find the energy scale below which
the low lying impurity doublet becomes substantially separated from
the excited symmetric orbital state.  Below this energy scale the
impurity complex can be regarded as a degenerate doublet, and the
system can be mapped to a 2CK model with only irrelevant perturbations
away from the intermediate coupling 2CK fixed point.  As a result, the
2CK fixed point is accessible at low temperatures for generic values
of the bare couplings provided that the conditions for an impurity
doublet obtain.

In order to demonstrate the existence of novel 2CK-like behavior at
low temperatures, one must do more than show that the RG flows at zero
temperature go to the 2CK fixed point. As shown by Affleck, Ludwig and
others, the singular low temperature behavior of various physical
quantities in the 2CK model is controlled by the {\em leading
  irrelevant operators} about the 2CK fixed point: i.e. by the flow
towards the fixed point.  We must thus show that the structure of the
irrelevant operators for the tunnelling impurity system is similar to
that of the 2CK model.  Indeed, we will show that the scaling
dimension of the leading irrelevant operators, which characterize the
form of the singularities in physical quantities at low temperatures,
is the same in the two cases. However the {\em number} of leading
irrelevant operators is larger for the tunnelling impurity system.
Analyzing the behavior close to the 2CK fixed point, we find {\em
  four} leading irrelevant operators, in contrast to the pure 2CK
model, which has only two, due to the fact that our problem does not
have the full symmetry of the conventional 2CK problem, $U(1)_{charge}
\times SU(2)_{orbital} \times SU(2)_{spin}$. We thus conclude that
since different physical quantities can have singularities controlled
by different combinations of the various leading irrelevant operators,
various amplitude ratios and scaling functions will be less universal
than in the 2CK problem, although the {\em form} of the singular low
temperature behavior will be the same.

In real systems, there will of course generally be some--hopefully
small--symmetry-breaking perturbations.  By studying the relevant
symmetry breaking perturbations, in particular strains, we conclude
that their effective strength will generally be {\em screened} by the
fluctuations at the 2CK fixed point, making their effects less
important than would be guessed by naively comparing energy scales.
However, for the experiments by Buhrman and
collaborators\cite{Ralph1,Ralph2,Buhrman} simple order-of-magnitude
estimates show that the strains have to be extremely small for the 2CK
fixed point to be approached.

Within the space of full triangular symmetry, we find that by tuning
the splitting between the impurity doublet and the singlet states,
(e.g. by pressure) there will be a phase transition at zero
temperature from the 2CK behavior to conventional Fermi liquid
behavior.  This transition is conjectured to be controlled by a fixed
point with $SU(3)$ orbital symmetry.

Impurity configurations with high symmetry, such as cubic, are also
discussed briefly, and found to be characterized, up to {\em
  additional} leading irrelevant operators, by the 2CK fixed point.

\subsection{Outline}

In the next Section we motivate and introduce the model we will be
using. In Section \ref{sec:IIIc3v}, we discuss the importance of the
impurity degeneracy and analyze three physically relevant situations,
for which the impurity ground state can indeed be two-fold degenerate.
The weak coupling analysis is performed in Section \ref{sec:IVc3v}, at
which the crossover from the 3-site problem to a degenerate 2-level
impurity system, as well as $C_{3v}$-symmetry breaking processes are
discussed. In Section \ref{sec:Vc3v}, the intermediate coupling regime
for the 2CK system is studied, while in Section \ref{sec:VIc3v}
conjectures about the behavior on and near the the critical manifold
separating 2CK from conventional Fermi liquid behavior are made.
Finally, we conclude in Section \ref{sec:VIIc3v} with the potential
relevance of the model to experimental situations and, in particular,
to recent experiments by Buhrman and
collaborators.\cite{Ralph1,Ralph2,Buhrman}

\section{Model}
\label{sec:IIc3v}

In this Section we introduce an effective Hamiltonian for the impurity
and conduction electron system, valid at some intermediate energy
scale substantially less than the conduction electron bandwidth. The
effects of higher energy electronic processes have already been taken
into account by introducing renormalized parameters in the effective
Hamiltonian. It is crucial that at this intermediate energy scale all
processes that are allowed by symmetry will be present, even if they
were absent in the ``bare'' Hamiltonian.  Therefore we have to include
all processes that are relevant (in the RG sense) at {\em either} weak
or intermediate coupling, since their amplitudes, albeit small
initially, can grow under the RG flows. [Other processes that are
irrelevant for both weak and intermediate coupling will end up
yielding at worst non-universal order one corrections to the various
quantities of interest.]

Since the impurity-electron interactions are predominantly local, the
electronic degrees of freedom which interact most strongly with the
impurity at low energies can be viewed as centered around the three
minimum energy positions of the impurity. Therefore for low enough
energies we only need to consider three electronic wavefunctions for
each energy and each spin. These can be represented as a symmetric
state $c^\dagger_{S\sigma\epsilon}$ with spin $\sigma = \uparrow
\downarrow$ and energy $\epsilon$, transforming according to the
trivial representation $\Gamma_1$ of $C_{3v}$, and an orbital doublet
\{$c^\dagger_{L\sigma\epsilon}$, $c^\dagger_{R\sigma\epsilon} $\}, in
the two dimensional representation, $\Gamma_3$. Neglecting weaker
interactions of the impurity with other electronic $ \Gamma_3
$-doublets, $\Gamma_1 $-singlets and antisymmetric $\Gamma_2
$-singlets will not change the behavior of the system. In an analogous
way the impurity states can be described by a $\Gamma_3$-doublet
$d^\dagger_L$, $d^\dagger_R$ and the symmetric state $d^\dagger_S$.

Before introducing the effective Hamiltonian ${\cal H}$, let us
summarize its symmetries. Impurity and electron numbers are separately
conserved, $d \rightarrow e^{i\phi} d$ and $c \rightarrow e^{i\theta}
c$.  In addition, in the absence of spin-orbit coupling or magnetic
fields, our system has spin-$SU(2)$ and time reversal symmetry. The
latter acts also on the orbital degrees of freedom: $i \rightarrow
-i$; $d_L \leftrightarrow d_R$; $c_L \leftrightarrow c_R$. Finally,
${\cal H}$ has the $C_{3v}$ symmetry, which includes exchange symmetry
around an axis ($c_L \leftrightarrow c_R$; $d_L \leftrightarrow d_R$)
and rotational symmetry by $\pm 2\pi/3$: $c_L \rightarrow
\exp\left[\pm 2\pi i/3 \right] c_L $; $\; \; c_R \rightarrow \exp\left[\mp
2\pi i/3 \right] c_R $; $\; \; c_S \leftrightarrow c_S$ and similarly for
the $d$'s.

The Hamiltonian ${\cal H}$ includes several types of terms:
\begin{equation}
  \label{Hchapterc3v}
  {\cal H}= {\cal H}_{el} + {\cal H}_{int} + {\cal H}_{hop} + {\cal
  H}_{mix}
\end{equation}
with ${\cal H}_{el}$  the electron kinetic energy in the absence of
the impurity
\begin{equation}
  \label{Ho}
  {\cal H}_0= \sum_\sigma \int \frac{d\epsilon}{2\pi} \epsilon
  \left[c^\dagger_{L\sigma\epsilon}c_{L\sigma\epsilon} +
   c^\dagger_{R\sigma\epsilon}c_{R\sigma\epsilon} +
   c^\dagger_{S\sigma\epsilon}c_{S\sigma\epsilon}\right],
\end{equation}
with the energy $\epsilon$  measured from the Fermi surface. Note
that for convenience we  consider a linear conduction band with a
cutoff of order the bandwidth $W$. Although we will
generally work in the $S$,$L$,$R$ representation, in order to see the
physical origin of various terms, it is convenient to first work
in a basis of wave-functions centered around the three equivalent
impurity sites, with
\begin{equation}
  \label{c_n}
  c_n = \frac{1}{\sqrt{3}} \left( c_S + e^{-i(n-1)\frac{2\pi}{3}} c_L +
 e^{i(n-1)\frac{2\pi}{3}} c_R \right)
\end{equation}
for $n=1,2,3$. If the impurity is at rest, then the only allowed one
electron terms are
\begin{equation}
  \label{V_1}
  V_1 \sum_{n \sigma} d^\dagger_n d_n c^\dagger_{n \sigma} c_{n \sigma}
\end{equation}
and
\begin{equation}
  \label{V_2}
  V_2 \sum_{i\neq j \neq n} \sum_\sigma  d^\dagger_n d_n c^\dagger_{i
  \sigma} c_{j \sigma}.
\end{equation}
The other such terms consistent with the symmetries can be rewritten,
using $\sum_{n} d^\dagger_n d_n =1$, in terms of these two and terms
which mix the electronic states centered at the different sites:
\begin{equation}
  \label{Hmix}
  {\cal H}_{mix} \propto \sum_{i\neq j} \sum_\sigma c^\dagger_{i
  \sigma} c_{j \sigma} = \sum_\sigma \left[ 2 c^\dagger_{S\sigma}
  c_{S\sigma} - c^\dagger_{L\sigma}
  c_{L\sigma} - c^\dagger_{R\sigma}
  c_{R\sigma} \right]
\end{equation}
which must, in any case, be present due to the non-orthogonality of
the $c_{1,2,3}$ states.\cite{MF2} [We ignore the unimportant term
$\sum_n c^\dagger_n c_n$].

The bare impurity hopping term can be written
\begin{equation}
  \label{Hhop0}
{\cal H}_{hop}^{(0)} = \frac{{\cal E}}{3} \sum_{i\neq j}
d^\dagger_i
d_j = \frac{{\cal E}}{3}
 \left( 2d^\dagger_S d_S - d^\dagger_R
 d_R - d^\dagger_L d_L \right)
\end{equation}
where the $d_{1,2,3}$ and $d_{L,R,S}$ are related as in Eq(\ref{c_n}).
The sign of ${\cal E}$ is of crucial importance.  In particular, if
all electron-impurity interaction terms are much smaller than ${\cal
  E}$, then the sign of ${\cal E}$ determines whether the impurity
ground state is an orbital singlet (${\cal E}< 0$) or a $\Gamma_3$
doublet (${\cal E} > 0$). At energy scales such that the
renormalized electronic bandwidth is smaller than $\left|{\cal E}
\right|$, in which case tunnelling events of the impurity from the
ground state configuration to excited states will be suppressed, the
low temperature behavior clearly depends crucially on the symmetry of
the impurity ground state in the absence of interactions.

Of course, as was noted in the
Introduction, naively the sign ${\cal E}$ should generally be
negative to ensure a non-degenerate ground state. However, situations
can exist for which the interaction of the impurity with atomic (high
energy) electrons conspires to make the effective low energy impurity
hopping matrix element ${\cal E}$ positive and hence favor the {\em
  doublet} impurity ground state. Some illustrative examples are
discussed in the next Section.

In addition to the bare impurity hopping, various ``electron
assisted'' impurity hopping terms (${\cal H}_{hop}^{(1)}$) can occur.
These may involve either terms of the form $d^\dagger_m d_n
c^\dagger_j c_j$ or $d^\dagger_m d_n c^\dagger_i c_j$ with $i\neq j$.
At the Fermi surface there are six independent such terms with
coefficients $A_1 \ldots A_6$. In the $L,R,S$ basis, these give rise
to several terms that are diagonal in this basis:
\begin{eqnarray}
  \label{Vdiag}
 {\cal V}_{diag} &=&  \pi Q_1 \sum_{\sigma} \left( c^\dagger_{L\sigma}
 c_{L\sigma} -
 c^\dagger_{R\sigma} c_{R\sigma} \right) \left( d^\dagger_L d_L -
 d^\dagger_R d_R \right) \\ \nonumber
 &+& \pi \frac{Q_2}{3} \sum_{\sigma} \left( 2c^\dagger_{S\sigma}
 c_{S\sigma}- c^\dagger_{R\sigma} c_{R\sigma} - c^\dagger_{L\sigma}
 c_{L\sigma} \right) \left( 2d^\dagger_S d_S - d^\dagger_R
 d_R - d^\dagger_L d_L \right) \\ \nonumber
&+& \pi Q_3 \left( c^\dagger_{S\sigma} c_{S\sigma} + c^\dagger_{L\sigma}
 c_{L\sigma} + c^\dagger_{R\sigma} c_{R\sigma} \right) \left(
 2d^\dagger_S d_S - d^\dagger_R
 d_R - d^\dagger_L d_L \right)
\end{eqnarray}
where $c^\dagger_{\mu\sigma}$, for $\mu= L, R, S$, are the ``local''
electron creation operators, represented as $c^\dagger_{\mu \sigma} \equiv
c^\dagger_{\mu \sigma} \left(x =0\right) $ where
\begin{equation}
  \label{c(x)}
 c^\dagger_{\mu\sigma}\left(x\right) \equiv \int \frac{d\epsilon}{2\pi}
  e^{-i\epsilon x} c^\dagger_{\mu\sigma\epsilon}
\end{equation}
are the incoming (outgoing) electronic $s$-wave states for $x > 0$
($x<0$).  The effect of the first two of these terms is to screen
transitions between the different states $L$, $R$ and $S$ of the
impurity.\cite{MF1} Their amplitudes $Q_1$, $Q_2$, $Q_3$ correspond to
linear combinations of various electron-assisted tunnelling processes
$(A_1 \ldots A_6)$ and, $Q_1$ and $Q_2$ will be assumed to be
positive. If $Q_1 <0$, then it will be seen later that the system
behaves like a ferromagnetic Kondo model [see discussion after
Eq(\ref{weakrgeqDelta3})]. In order for these important terms to be of
appreciable magnitude (in Section \ref{sec:Vc3v} we will allow $Q_1$
to be of order unity), we focus on impurities with relatively fast
(renormalized) tunnelling rates.

The interaction terms $V_1$ and $V_2$ give rise to a number of
processes in the $L,R,S$ basis which change the state of the impurity;
to be consistent with $C_{3v}$ symmetry the total orbital ``angular
momentum''-- with $L=-1$, $R=+1$ and $S=0$ -- of each of these terms
must be zero (mod 3). The electron assisted hopping terms $(A_1 \ldots
A_6)$ also contribute to ${\cal V}_{off-diag}$. We have
\begin{eqnarray}
  \label{Voffdiag}
  {\cal V}_{off-diag} &=&  2\pi\Delta_{1}
  \sum_{\sigma} \left(d^\dagger_L d_R c^\dagger_{R\sigma} c_{L\sigma}
  + h.c. \right)
  \\ \nonumber
  &+&  2\pi\Delta_{2}
  \sum_{\sigma} \left(d^\dagger_L d_S c^\dagger_{S\sigma} c_{L\sigma}
  + d^\dagger_R d_S c^\dagger_{S\sigma} c_{R\sigma} + h.c. \right)
  \\ \nonumber
  &+& 2\pi\Delta_{3}
  \sum_{\sigma} \left(d^\dagger_L d_R c^\dagger_{L\sigma} c_{S\sigma}
  + d^\dagger_L d_R c^\dagger_{S\sigma} c_{R\sigma}
  + h.c. \right) \\ \nonumber
  &+&  2\pi\Delta_{4}
  \sum_{\sigma} \left(d^\dagger_L d_S c^\dagger_{R\sigma} c_{S\sigma}
  + d^\dagger_R d_S c^\dagger_{L\sigma} c_{S\sigma} + h.c. \right) \\
  \nonumber
  &+&  2\pi\Delta_{5}
  \sum_{\sigma} \left(d^\dagger_L d_S c^\dagger_{L\sigma} c_{R\sigma}
  + d^\dagger_R d_S c^\dagger_{R\sigma} c_{L\sigma} + h.c. \right)
\end{eqnarray}
 In the absence of impurity motion,  ${\cal V}_{diag}$ is zero, but ${\cal
V}_{off-diag}$ has contributions from both
electron-impurity interactions and electron-assisted hopping and is
thus still non-zero.

Of all the terms we have so far, $\Delta_3$ and $\Delta_5$ are the
only processes, that {\em break} the $O(2)$-symmetry in the $\Gamma_3
$-manifold (which, using different notation, was discussed in
Ref. \cite{MF2} in the context of the two-site tunnelling
impurity). However, these terms are clearly $C_{3v}$ invariant and
their effects will be discussed in later Sections.

The terms we have included so far can be conveniently grouped into
\begin{equation}
  \label{H0}
  {\cal H}_0 = {\cal H}_{el} + {\cal H}_{hop}^{(0)}
\end{equation}
and the ${\cal V}_{diag}$ and ${\cal V}_{off-diag}$ combinations
exhibited above. The electron mixing term, as it is diagonal in the
$L,R,S$ basis,will not play a substantial role (in contrast with the
two-site case of Ref. \cite{MF2}). In addition to the various
impurity one-electron coupling terms, the impurity can also affect
electron-electron interactions, and two-electron assisted hopping can
 take place. Although these terms  are strongly irrelevant for weak
coupling as discussed in Ref. \cite{MF2}, the latter can play an
important role near the intermediate coupling 2CK fixed point; we
denote them as ${\cal H}_{hop}^{(2)}$. Putting all the important terms
together, we  primarily study the
effective Hamiltonian
\begin{equation}
  \label{Heff}
  {\cal H} = {\cal H}_0 + {\cal V}_{diag} + {\cal V}_{off-diag} + {\cal
  H}_{hop}^{(2)}.
\end{equation}

At this point, one can immediately guess which terms will play
dominant roles at very low energy scales where, with ${\cal E}$
positive, the higher energy symmetric impurity state $d^\dagger_S$
will be strongly suppressed. In this limit, the most important terms
will be $Q_1$ which is analogous to the $J_z$ coupling in the Kondo
system, and $\Delta_1$ which is analogous to the $J_\perp$
coupling. The only other one-electron term which does not involve
$d_S$ is $\Delta_3$ which involves the $c_S$ electrons and breaks the
$O(2)$ symmetry of the $L,R$ subspace.

In general there can also be  $C_{3v}$-symmetry breaking
processes, but, by assumption, these will be small. Nonetheless we
will need to investigate their role. The most
important terms  correspond to bare impurity transitions from state
$L$  to $R$, such as
\begin{equation}
  \label{C3vbreakingterms}
   \left( M d^\dagger_L d_R + M^* d^\dagger_R
  d_L\right),
\end{equation}
 or from $L$ and $R$ to $S$, such as $d^\dagger_S \left( d_L +
d_R \right) + \left(d^\dagger_L + d^\dagger_R \right) d_S$. [Note that
$d^\dagger_L d_L - d^\dagger_R d_R$ and $d^\dagger_S \left( d_L - d_R
\right) + (d^\dagger_L - d^\dagger_R) d_S$ cannot appear if the system
is time-reversal invariant]. These will originate from various
effects, including strains, other nearby static impurities etc. Of the
above processes, the one corresponding to $M$ will be the most
important to our discussion, because it will persist even on energy
scales below which ${\cal E}$ becomes larger than the renormalized
bandwidth and transitions to the $S$-state are effectively quenched.
The real part of $M$ will be a linear combination of two independent
bare impurity terms that are even under parity $d^\dagger_L
\leftrightarrow d^\dagger_R$, i.e. an energy bias term $2 d^\dagger_1
d_1 - d^\dagger_2 d_2 - d^\dagger_3 d_3$ and an asymmetry in the
hopping amplitudes $d^\dagger_2 d_3 + d^\dagger_3 d_2$, while in the
imaginary part of $M$ similar terms will appear which are odd under
parity. Thus, two independent parameters have to be tuned for $|M|$ to
vanish. Finally, note that asymmetries in the electronic states due to
other impurities etc., can generate $M$ under renormalization.

\section{Jahn-Teller effect, Berry's phase and Excited Impurity
  Configurations: Effects on Impurity Degeneracy }
\label{sec:IIIc3v}

As already mentioned, the degeneracy of the lowest lying effective
impurity state at intermediate energy scales is crucial to our
analysis. The generic case of a non-degenerate impurity ground state,
corresponding to ${\cal E} < 0$, leads to uninteresting impurity
screening and hence to ordinary Fermi-liquid behavior at low
temperatures. Instead, a doubly degenerate $\Gamma_3$ orbital impurity
state will be our focus, and in this Section we will show how the
necessary condition, ${\cal E} > 0$, is possible. First, we will
analyze in two model cases the effects on the impurity motion of high
energy atomic or valence electrons in solids, ignoring coupling to the
conduction electrons. The impurity will be assumed to move slowly on
the scale of the atomic electronic time scales. We can thus consider
the adiabatic evolution of the electronic ground state as the impurity
moves. In the second part of this Section we will demonstrate how,
when we include the conduction electrons, the existence of
intermediate excited impurity configurations between the low lying
states $d^\dagger_1$, $d^\dagger_2$, $d^\dagger_3$, can serve as a
mechanism for changing the sign of the effective ${\cal E}$ at lower
energies.

\subsection{Effects of atomic electrons on impurity degeneracy}

We focus first on an orbital- and spin-singlet interstitial impurity,
tunnelling between three equivalent positions--1, 2 and 3--on the
corners of an equilateral triangle.  The underlying physical reason
for the desired effect on the sign of ${\cal E}$ is the appearance
of degeneracies in the states of the surrounding medium (electrons and
neighboring nuclei) when the impurity is located at a position of
special symmetry--in our case, the center of the triangle.  Even if
the impurity is forbidden energetically from occupying this central
symmetric position, e.g. by the presence of a host lattice atom, the
effect is manifested in the behavior of the wavefunctions of the
system when the impurity is away from the center.

The simplest example is an atomic electronic state on the impurity,
which, with the impurity located in the center of the triangle, is
doubly degenerate, i.e. an $L, R$ doublet. When the impurity is at,
say, site 1, the degeneracy is broken. However, by adiabatically moving
the impurity once around the triangle and back to site 1, the electron
returns to its original state, but with its wavefunction multiplied by
a factor of $-1$. The extra phase acquired by the electronic state is
just Berry's phase;\cite{Berry} it will change the sign of the
effective impurity-plus-electron tunnelling matrix elements.

With one electron in an appropriate energy level on the impurity, this
effect can occur but the impurity will be magnetic. What is needed
instead is an {\em orbital} (non-magnetic) electronic doublet on the
impurity when it is at the center of the triangle.

This is a situation which is a classical example of the
Jahn-Teller\cite{Jahn-Teller} effect and was also proposed recently in
this context by Gogolin.\cite{Gogolin0} Due to the coupling of the
electronic doublet of the impurity and the neighboring atoms, the
triangle-center position of the impurity is unstable to displacements of the
impurity from its equilibrium position: this linearly splits the
electronic degeneracy, but only costs elastic energy that is
quadratic in the distortion. A massive impurity will be displaced to
one of three minimum energy equilibrium positions on the corners of an
equilateral triangle. But, since the impurity wavefunctions at these
three positions will generally have a finite overlap, the impurity can
tunnel between them. The phases of the tunnelling matrix elements add
up to $\pi$ on tunnelling around the triangle, yielding an extra minus
sign from the Berry's phase.  This ensures that the tunnelling
$\Gamma_3$-doublet will be the ground state of the impurity complex.
This restoration of the degeneracy via tunnelling of the impurity is a
dynamic Jahn-Teller effect--the triangular symmetry is not broken in
the ground state and the original electronic doublet has merely been
modified into an impurity orbital doublet.

Unfortunately, orbital electronic doublet ground states at high
symmetry positions of an impurity are not too common. Instead, in many
cases the minimum energy configuration has magnetic character, due to
Hund's rules splittings. Nevertheless, for large enough local lattice
distortions due to the impurity, the effective tunnelling impurity
ground state can still be a doublet. To explicitly demonstrate this
effect, we consider a simple situation with four atomic electrons on
the impurity, perturbed by the surrounding medium, with spin-orbit
interactions neglected. For simplicity we consider only three electronic
levels which, with the impurity in the central point with triangular
symmetry, will have the symmetry of the $L$, $R$ and $S$ states,
corresponding to angular momentum of $m=\pm 1, 0$ (mod 3)
respectively. We denote the creation operators for these
$a^\dagger_{L,R,S}$. If the environment is such as to make the atomic
one-electron $S$ state the lowest by energy $\epsilon_a$, two electrons
will fill this and the other two will go in the degenerate $L$, $R$
states. The simplest situation is with effective electron-electron
interactions smaller than the splitting $\epsilon_a$. In this case,
Hund's rules will give rise to ferromagnetic exchange, $J$, between
the two $L$ or $R$ electrons, resulting in a spin triplet $ ^3S$
ground state, with orbital angular momentum (mod 3), m=0, a doublet
pair $ ^1E$ of spin singlet, $m=\pm 1$ states ($| L_\uparrow
L_\downarrow>$ and $| R_\uparrow R_\downarrow>$), and a higher energy
symmetric singlet, $ ^1 S$, excited state. Coupling to the
one-electron $S$ states will mix, e.g., $| S_\uparrow R_\downarrow -
S_\downarrow R_\uparrow>$ in with the $| L_\uparrow L_\downarrow>$
state with some amplitude $K$. These processes can be represented
by a Hamiltonian for the impurity at the triangle center,
\begin{eqnarray}
  \label{Hundterms}
  {\cal H}_{center}&&\! \! \! = -\epsilon_a \sum_{\sigma = \uparrow,
    \downarrow} a^\dagger_{S\sigma} a_{S\sigma} + J \left(
  a^\dagger_{L\uparrow} a_{R\uparrow} a^\dagger_{R\downarrow}
  a_{L\downarrow} + h.c. \right) \\ \nonumber
&& \! \! \!+ K \left(
  a^\dagger_{S\uparrow} a_{L\uparrow} a^\dagger_{R\downarrow}
  a_{L\downarrow} + a^\dagger_{S\uparrow} a_{R\uparrow}
  a^\dagger_{L\downarrow} a_{R\downarrow} + a^\dagger_{R\uparrow}
  a_{L\uparrow} a^\dagger_{S\downarrow} a_{L\downarrow} +
  a^\dagger_{L\uparrow} a_{R\uparrow} a^\dagger_{S\downarrow}
  a_{R\downarrow} + h.c.\right)
\end{eqnarray}
[Note that if all four electrons are
considered together, with the effective interactions being larger than the
one-electron splittings caused by the environment, Hund's rules would still
result in an $m=0$ triplet ground state, but with the lowest excited
states also being spin-triplets.]

\begin{figure}[htpb]
    \begin{center}
      \leavevmode
      \epsfxsize=3truein
      \epsfbox{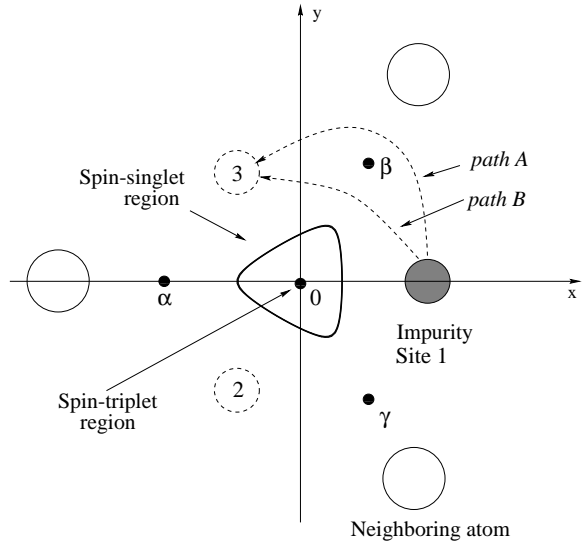}
    \end{center}
    \caption[Representation of impurity paths]{Schematic
      representation of the minimum energy impurity positions (1,2,3)
      in the environment of the neighboring atoms (open circles). The
      solid triangle is the boundary between the regions in which the
      electronic ground state is a spin-triplet (interior) and a
      spin-singlet (exterior). Points 0, $\alpha$, $\beta$, $\gamma$
      represent positions of the impurity at which the two lowest
      lying electronic spin-singlet levels are degenerate. At point 0
      these two states form an orbital $ ^1E$ doublet [see
      Fig.\ref{fig:E(x)} and Section \ref{sec:IIIc3v}.1]. This figure
      corresponds to large $\epsilon_a$, with the parameters defined
      in Section \ref{sec:IIIc3v}.1 $\epsilon_a = 33.3 J = 10 K = 20
      \delta_1 = 10 \delta_2$ and units of $x$ and $y$, such that
      $\alpha$, $\beta$ and $\gamma$ are at distance approximately 0.6
      from $0$, see Fig.\ref{fig:E(x)}.  Depending on the choice of
      these parameters, the crossing points $\alpha$, $\beta$ and
      $\gamma$ may lie inside or outside of the triangle.  Two types
      of lowest action tunnelling paths ``A'' and ``B'' are shown and
      discussed in Section \ref{sec:IIIc3v}.1.}
     \label{fig:Berrypaths}
\end{figure}

Motion of the impurity away from the  triangle center by a
displacement $(x,y)$ with $-x$ in the direction of one of the three
neighboring atoms, will give rise to mixing of the one-electron
states via terms of the form
\begin{eqnarray}
  \label{levelsplitting}
 {\cal H}_{split} = &-&\delta_1 \sum_{\sigma = \uparrow, \downarrow}
 \left[ a^\dagger_{L\sigma}
 a_{R\sigma} (x+ iy) + h.c. \right] \\ \nonumber
&-&\delta_2 \sum_{\sigma = \uparrow, \downarrow} \left[
 a^\dagger_{R\sigma} a_{S\sigma}  (x+ iy) + a^\dagger_{L\sigma}
 a_{S\sigma} (x-iy)  +
h.c. \right]
\end{eqnarray}
These will give rise to splittings of the doublet excited state that
are {\em linear}
in the displacement; this is just what-would-be the Jahn-Teller
splitting if the doublet had been the ground state. If the
displacement is directly towards or away from one of the triangle
corner atoms, say, with  $y=0$, the states can still be categorized as even
or odd under reflection through that atom. The even state will also
mix with the $m=0$ spin singlet excited state.

We are interested in distortions large enough that the ground state is
no longer the spin triplet, but one of the even or odd singlet states.
It can readily be seen in simple models, like the one above, that this
can occur, at least in principle. There will be then, for a classical
impurity, three equivalent spin-singlet ground states, with
displacements $(R, 0)$, $(-R/2, \pm \sqrt{3}R/2)$.  In general, the
lowest action tunnelling path between these positions will go around
the excited state doublet degeneracy in the center of the triangle,
and hence pick up a minus sign from Berry's phase.

\begin{figure}[htpb]
    \begin{center}
      \leavevmode
      \epsfxsize=3truein
      \epsfbox{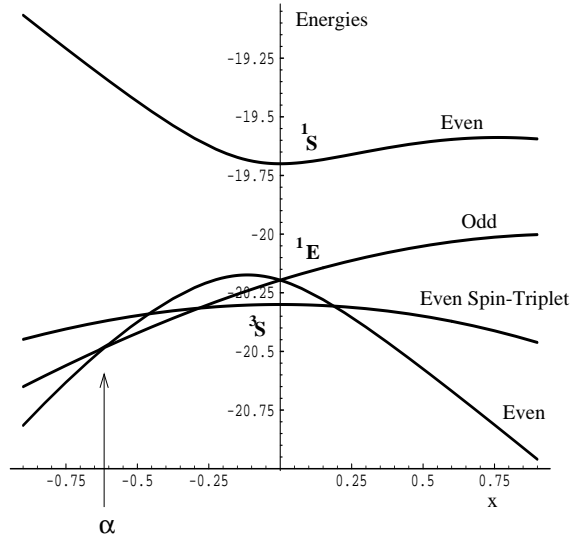}
    \end{center}
    \caption[Dependence of energies of low lying atomic electron
    states as a function of impurity position]{Energies of the low
      lying atomic electron states, discussed in Section
      \ref{sec:IIIc3v}.1, as a function of the impurity displacement
      $x$ along the $y=0$ axis [see Fig.\ref{fig:Berrypaths}]. This
      plot corresponds to $\epsilon_a = 33.3 J = 10 K = 20 \delta_1 = 10
      \delta_2$, the same as in Fig.\ref{fig:Berrypaths}. When $x=0$ the
      impurity is located at the triangle center and the electronic
      orbital doublet $ ^1 E$ is between the $ ^3S$ spin-triplet,
      orbital-singlet and the $ ^1S$ spin- and orbital- singlet
      states. When $x\neq 0$ (but $y=0$)the orbital doublet is split
      linearly and all the states are now classified as even or odd
      under reflection through the $x$-axis.  Due to repulsion between
      the spin-singlet even states, for $x<0$ the lower even state
      crosses the odd one at point $\alpha$ as also shown in
      Fig.\ref{fig:Berrypaths}.}
     \label{fig:E(x)}
\end{figure}

Unfortunately,
there may also be other degeneracies in the spin singlet sector that
can play a role. In particular, level repulsion between the even
singlet states along the line $y=0$ (and two equivalent lines), can
result in a second crossing between the even and odd singlet states.
It should be stressed that these extra level crossings
$\alpha,\beta,\gamma$ [see Fig.\ref{fig:Berrypaths}] are {\em distinct
  points}, occuring only when the impurity lies on one of the three
lines of symmetry, so that the two low-lying singlet states have
different (even-odd) symmetry and hence can cross, see
Fig.\ref{fig:E(x)}. In particular, if the modification of the
effective interactions by the distortion is small, (e.g. for
$\epsilon_a \gg J, K$ with no modification of the interactions that
give rise to $J$ and $K$ due to the distortion), the main effect of
the distortion will be the repulsion between the even singlet states;
this situation is shown schematically in Fig.\ref{fig:E(x)}. In this
case the degeneracies and region of spin singlet or triplet ground
states are shown in Fig.\ref{fig:Berrypaths}. If the lowest action
tunnelling path is like that denoted by ``A'' in
Fig.\ref{fig:Berrypaths}, {\em four} degeneracies in the spin singlet
sector will be encircled as the impurity moves around the triangle,
resulting in four minus signs and hence a conventional symmetric
orbital ground state of the tunnelling impurity, i.e. ${\cal E} <0$.
On the other hand, if the impurity tunnels via a path like that shown
as ``B'' in Fig.\ref{fig:Berrypaths}, only the triangle center
degeneracy will be encircled and the tunnelling ground state will be
the impurity $L,R$ doublet. Either possibility can occur. It should
 be mentioned that if the repulsion between the electronic states
($J$, $K$) is also modified by the distortion, then the three extra
degenerate points can be made to disappear in simple models, thereby
leaving only the triangle center degeneracy and only paths of type
``B'' possible, hence making the impurity ground state a doublet.

Note that spin flip effects can complicate matters somewhat, but will
not change the overall conclusion.

We have shown in this Section that the existence of an electronic
degeneracy in either the ground state or some particular excited state
when the impurity is located in a special, albeit energetically
unfavorable, point in space (the triangle center) can be responsible
for the restoration of that degeneracy for the combined
impurity-plus-environment system. This only occurs in some regimes of
atomic electron interactions and couplings to the environment.
Clearly, a complicated calculation would be required to determine the
amplitudes of the various processes reasonably accurately in an actual
physical example. Furthermore, the above analysis is valid only when
the impurity moves slowly compared to the atomic electron time scales.
One should note that the above assumption, although intuitively
obvious in many cases (especially away from electronic level
crossings) is still controversial.\cite{Hauge} However the above
discussion demonstrates that, at least in principle, the appearance of
a Berry phase, and therefore an orbitally degenerate impurity ground
state, is indeed possible.

\subsection{Effect of an excited impurity configuration}

In realistic situations, impurity tunnelling rates will probably be
appreciable if the barriers between the three impurity positions are
relatively small. Therefore the excited energies of intermediate
impurity configurations that are comparable to these barriers, will be
substantially lower than the bandwidth. As a result, a considerable
fraction of conduction electron excitations will have energies large
enough to allow transitions between the low lying and these excited
impurity states. However, since the electronic excitation spectrum is
continuous in the thermodynamic limit, the impurity motion cannot be
treated adiabatically for these energies, as it was for the atomic
electron states discussed above. Instead, the conduction electron
assisted tunnelling of the impurity between the low lying and these
excited states can be analyzed systematically by RG scaling. Here we
show how the existence of an excited impurity configuration can lead
to a positive effective ${\cal E}$ at low energies via coupling to
conduction electrons.

\begin{figure}[htpb]
    \begin{center}
      \leavevmode
      \epsfxsize=3truein
      \epsfbox{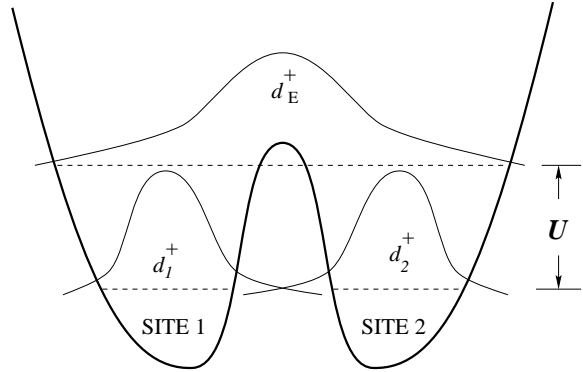}
    \end{center}
    \caption[Effect of excited impurity states]{Schematic illustration
      of a double well potential for an impurity with low lying states
      $d^\dagger_1$, $d^\dagger_2$ (and their representative
      wavepackets localized around sites 1 and 2, respectively). The
      wavefunction of the excited impurity state $d^\dagger_E$ at an
      energy $U$ higher than $d^\dagger_1$, $d^\dagger_2$ [see Section
      \ref{sec:IIIc3v}.2], as shown, has significant amplitude
      in both wells. }
     \label{fig:excitedstate}
\end{figure}

For simplicity, we only consider two sites, with
the low lying impurity states at each site being $d^\dagger_1$,
$d^\dagger_2$.  Let $d^\dagger_E$ denote an intermediate excited
impurity state, with extra energy $U$, that is large compared to the
bare impurity tunnelling matrix element between sites 1 and 2, but
small compared to the bandwidth $W$ ($U/W\ll 1$), see
Fig.\ref{fig:excitedstate}. In this regime we can perform a two-stage
renormalization procedure: Initially we include the excited state,
allowing electron assisted tunnelling events between states 1, 2 and
$E$, these may have tunnelling amplitudes $t_1$ much greater than
${\cal E}$.  Then, when the renormalized bandwidth becomes of the
order of $U$ or smaller, excitations of the impurity to state $E$ will
become negligible, in which case we can drop the excited state from
the effective Hamiltonian. Thus the effect of the impurity excited
state can be described by the following terms
\begin{eqnarray}
  \label{excitedstateH}
  \Delta {\cal H} &&= U d^\dagger_E d_E +  y \sum_\sigma \left(
  c^\dagger_{1\sigma}  c_{2\sigma}
  + c^\dagger_{2\sigma}  c_{1\sigma} \right)
  \nonumber \\
  &&+  2 \pi t_1 \sum_\sigma \left(  d^\dagger_E d_1
  c^\dagger_{1\sigma} c_{E\sigma} + d^\dagger_E d_2
  c^\dagger_{2\sigma} c_{E\sigma}  + h.c. \right)
\end{eqnarray}
where $c_{E\sigma}^\dagger$ is an extra local electronic state, which
is even under the exchange of site indexes 1 and 2 and $y$ is the {\em
  average} mixing amplitude of conduction electron states 1 and 2 in
the presence of the impurity. Note that in the absence of the
impurity, $y$ (together with the other terms in $\Delta {\cal H}$)
will vanish, restoring the orthogonality of states 1 and 2. In
general, both $t_1$ and $y$ are energy dependent.\cite{MF1} For
simplicity, we only consider their constant part, evaluated at the
Fermi energy. Energy dependent deviations will introduce corrections
but will not qualitatively change the behavior described
here.\cite{MF1} A crucial observation is that $y$ can be positive or
negative due to two effects. First, as it was seen in Ref. \cite{MF1},
since $c^\dagger_{1\sigma}$, $c^\dagger_{2\sigma}$ involve electronic
excitations in the conduction band, $y$ is an oscillatory function of
$k_F R$, where $k_F$ and $R$ are the Fermi wavevector and the
intersite distance, respectively. Roughly, this oscillatory behavior
corresponds to the amplitude of an electron at the Fermi energy
tunnelling between sites 1 and 2. Second, since it
is the presence of the impurity that causes the mixing of the
otherwise orthogonal states 1 and 2, $y$ should also depend on the
impurity-electron coupling. As a result, its sign should also depend
on the sign of the coupling.\cite{MF1}

Neglecting all other terms involving the impurity, in this regime the
RG equation of ${\cal E}$  normalized by
the conduction bandwidth $W$ is
\begin{equation}
    \label{RGE0}
    \frac{d{\cal E}}{dl}= {\cal E} + 6 y t^2,
\end{equation}
while $y$ and $t_1$ are marginal (for weak electron-impurity
interaction).  The second term ceases to contribute to the
renormalization of ${\cal E}$ once the renormalized bandwidth becomes
of the order of $U$. At that scale $e^{{\tilde l}} = W/U$, ${\cal
  E}({\tilde l}) \approx ({\cal E} + 6 y t_1^2)W/U$ so that (after
reinstating the renormalized bandwidth factors) the
effective renormalized ${\tilde {\cal E}}$ that should appear in the lower
energy effective Hamiltonian, Eq(\ref{Hhop0}), is
\begin{equation}
  \label{Eren}
    {\tilde {\cal E}} \approx {\cal E} + 6y t_1^2 W,
\end{equation}
independent of $U$.  Below the energy scale $U$, the impurity is
limited to hops between the low lying states $d^\dagger_1$,
$d^\dagger_2$ (and $d^\dagger_3$), with an effective hopping ${\tilde
  {\cal E}}$. Note that since the $d^\dagger_E$ state has appreciable
amplitude in both wells, $t_1^2 W$ may be larger than ${\cal E}$.
Therefore $y t_1^2 W$ may be comparable to ${\cal E}$, and, since the
sign of $y$ is arbitrary, ${\tilde {\cal E}}$ may be positive or
negative, whatever ${\cal E}$ may have been originally. [Note that
there will also be contributions to ${\tilde{\cal E}}$ from hops
$1\rightarrow E \rightarrow 2$ {\em not} involving electrons, but
these need not dominate over the electron-assisted process included here.
The effects of an excited impurity configuration were first considered
by Zar\'{a}nd and Zawadowski\cite{Zarand} in the context of an
impurity tunnelling between two sites, in order to show that the
one-electron assisted tunnelling between the low-lying levels (which,
for simplicity, we did not consider here) can be enhanced. However,
they did {\em not} consider the equally strong enhancement of the tunnelling
rate of the impurity {\em alone}, which we analyzed above.]

In summary, we have presented three situations which can
result in an impurity $L,R$ doublet ground state at low energies. The
first two (Section \ref{sec:IIIc3v}.1) involved the effect of
localized atomic electrons with energies far from the Fermi level. In
the third case (Section \ref{sec:IIIc3v}.1) we saw that, closer to the
Fermi level, conduction electrons in the presence of excited impurity
configurations can renormalize the effective bare tunnelling element
${\cal E}$, so that it can take either sign. After all these effects have
been taken into account, at intermediate energy scales the
coupling of the remaining conduction electron degrees of freedom to
the impurity can be represented by the {\em effective} renormalized
coupling terms introduced in Section \ref{sec:IIc3v}; these will now
be analyzed.

\section{Weak tunnelling crossover}
\label{sec:IVc3v}

In most physical situations, it is likely that the bare impurity
tunnelling ${\cal E}$ and the electron assisted tunnelling processes
will be small. If the impurity-position-dependent electron-impurity
interactions $V_1$ and $V_2$ are also small, then there will be a
simple crossover at an energy scale of order ${\cal E}$. At lower
energy scales, the processes that involve $d_S$ will be strongly
suppressed; those remaining will be $\Delta_1$, $Q_1$ and $\Delta_3$.
By this crossover scale, the various $Q_j$ and $\Delta_j$, as they are
all marginal for weak coupling, will only have been modified from
their original values by amounts quadratic in the various couplings;
${\tilde{\cal E}}$ will have been renormalized similarly in addition to its
simple growth under renormalization with eigenvalue one. For weak
coupling, two-electron terms, like those in ${\cal H}^{(2)}_{hop}$,
will be irrelevant with eigenvalue $-1$ and hence will be negligible
at the crossover scale.

Beyond this crossover scale for weak interactions, the couplings other
than $\Delta_1$, $\Delta_3$ and $Q_1$ can be ignored (at least for
now), but these will renormalize each other. With all three of these
small the flow equations in this regime are
\begin{eqnarray}
  \label{weakrgeqnsQ1Delta1}
  \frac{d Q_1}{d l} &=& 2 \Delta_1^2 \\ \nonumber
\frac{d \Delta_1}{d l} &=& 2 Q_1 \Delta_1
\end{eqnarray}
(reflecting the approximate $SU(2)$ symmetry for $\Delta_1 = Q_1$) and
\begin{equation}
  \label{weakrgeqDelta3}
  \frac{d \Delta_3}{d l} = - Q_1 \Delta_3.
\end{equation}
If $Q_1 < - \left|\Delta_1 \right|$, then the system is like a
ferromagnetic Kondo problem so that $\Delta_1$ flows to zero and the
ground state will be an impurity orbital doublet. We thus restrict
consideration to $Q_1 > - \left|\Delta_1 \right|$. Since $Q_1$
involves impurity hopping but $\Delta_1$ does not, we expect $\left|
Q_1 \right| <  \left|\Delta_1 \right|$, although this need not be the
case.

We thus see that $\Delta_3(l)$ will tend to decrease in the
interesting regime, while $Q_1$ and $\Delta_1$ increase. The Kondo
temperature $T_K$, below which the intermediate coupling behavior will
obtain, is determined by the energy scale at which $\Delta_1(l)$
becomes of order unity, yielding
\begin{equation}
  \label{TK}
  T_K \sim W \exp\left[ -\frac{\kappa}{\left| \Delta_1 \right|}\right]
\end{equation}
with $W$ the electron bandwidth and $\kappa$ depending on the ratio of
$Q_1$ and $\Delta_1$ initially, or more precisely, at the crossover
scale at which ${\cal E}(l)$ is of the order the renormalized
bandwidth; specifically for $\left| Q_1\right| < \left| \Delta_1
\right|$,
\begin{equation}
  \label{kappa}
  \kappa = \frac{1}{ \sqrt{ 1- \frac{Q_1^2}{\Delta_1^2}}} \arctan
  \sqrt{ \frac{1- \frac{Q_1}{\Delta_1}}{1+ \frac{Q_1}{\Delta_1}}}
\end{equation}
which is $\pi/4$ in the small $Q_1$ limit.

If the interactions of electrons with the impurity are reasonably
strong, i.e $V_1$ and/or $V_2$ are not small, then the RG flow away
from weak tunnelling is more complicated. In particular, the weak
tunnelling eigenvalue of the impurity hopping, $\lambda_{{\cal
      E}}$ is modified to
\begin{equation}
  \label{lambdaDelta0}
  \lambda_{{\cal E}} = 1 - \frac{1}{6\pi^2} \left(V_1^2  +2
  V_2^2 \right)
\end{equation}
 by orthogonality catastrophe effects. This expression is
valid for $V_1$ and $V_2$ of order unity if they are replaced by the
associated phase shift-like quantities, see Ref. \cite{MF1}. The
renormalization of the various electron assisted tunnelling terms
$A_1, \ldots A_6$ will also be modified by the impurity-electron
interactions, as discussed in Ref.\cite{MF1}. Indeed, for
sufficiently large $V_1$ and/or $V_2$ some of these can be more
relevant than ${\cal E}$, although in this case the feedback of these
into ${\cal E}$ will become important. Likewise, ${\cal E}$ combined
with ${\cal H}_{mix}$ will generate the electron-assisted hopping
terms if these do not already exist.

Thus for strong interactions, the weak tunnelling crossover will be
rather complicated with several processes potentially playing
significant roles. But the coupling of primary importance for the
Kondo-like physics, $\Delta_1$, can itself be quite large initially;
in this case the Kondo temperature will essentially be when ${\cal E}$
becomes of order the renormalized bandwidth i.e.
\begin{equation}
  \label{TKforlargeV}
  T_K \sim W \left({\cal E}/W\right)^{1/\lambda_{{\cal E}}}
\end{equation}
with $\lambda_{{\cal E}} \leq 1$. Note that initially $Q_1$ will not
be generated by the $V_i$ terms, although they will affect its
renormalization, but once the effective ${\cal E}$ becomes
appreciable, $Q_1$ will be generated by $\Delta_1$ as in
Eq(\ref{weakrgeqnsQ1Delta1}) and then rapidly become of order unity.

\subsection{Symmetry Breaking Processes}

The dominant $C_{3v}$ symmetry breaking process is given by
Eq(\ref{C3vbreakingterms}). For weak coupling $M$ is relevant with
 eigenvalue one. Beyond the crossover energy scale
at which ${\cal E}(l)$ becomes large, $M$ can be renormalized
non-trivially by $Q_1$ and $\Delta_1$. If $\Delta_1$ is large, there
will not be a substantial regime in which this will occur and we can
simply use the magnitude of $M$ at this crossover as an input to the
flows in the interesting intermediate coupling regime analyzed in the
next Section. On the other hand if $\Delta_1$ (and $Q_1$) is small,
then there will be a regime of non-trivial renormalization of $M$,
with corrections to its eigenvalue of order $\Delta_1^2$ and $Q_1^2$.
But since the logarithmic range of scales over which this will obtain
is of order $l \sim 1/\Delta_1$, the net effect of this will be small.

Thus in both regimes, the weak tunnelling crossover will not
appreciably alter the effective magnitude of the dominant symmetry
breaking perturbations.

\section{Intermediate Coupling 2CK Fixed Point}
\label{sec:Vc3v}

We now analyze the intermediate coupling regime and show how
two-channel Kondo-like behavior arises.  We work in the regime where
the gap between the symmetric impurity state and the lower lying
$\Gamma_3 $-doublet is much larger than the renormalized electronic
bandwidth, i.e. at energy scales much less than the Kondo temperature
$T_K$. In this case the impurity is essentially frozen in its doublet
ground state and we can therefore impose the constraint
\begin{equation}
  \label{2levelconstraint}
  d^\dagger_L d_L + d^\dagger_R d_R =1
\end{equation}
and limit the analysis to processes involving only these two
states. In this Section we first identify the 2CK fixed point, which is
located at intermediate  values of the important coupling parameters
$Q_1$ and $\Delta_1$ and therefore
cannot be accessed using a weak coupling analysis. We then show
that it is stable when $C_{3v}$ is an exact symmetry. Finally we  find
the set of leading irrelevant operators which will control  the low
temperature behavior of the system.

It is convenient to introduce pseudo-spin operators for the impurity
doublet
\begin{equation}
  \label{defsigmaz}
  \sigma_z = d^\dagger_L d_L -d^\dagger_R d_R
\end{equation}
and
\begin{eqnarray}
  \label{defsigmapm}
  \sigma_+ &=& d^\dagger_L d_R \\ \nonumber
 \sigma_- &=& d^\dagger_R d_L
\end{eqnarray}
In order to proceed with our analysis we bosonize the electronic
degrees of freedom by introducing bosonic fields $\Phi_{\mu \sigma}
\left(x \right)$ for $\mu=L,R,S$ and $\sigma= \uparrow, \downarrow$ in
  the following way:
\begin{eqnarray}
  \label{cmusbosonization}
  c_{L\uparrow} \left( x\right) &=& \frac{1}{\sqrt{2\pi\tau_c}}
  e^{i\Phi_{L\uparrow}\left(x\right)} e^{i\pi N_{S\uparrow}}
  \\  \nonumber
  c_{R\uparrow}\left(x\right) &=& \frac{1}{\sqrt{2\pi\tau_c}}
   e^{i\Phi_{R\uparrow}\left( x\right) } e^{i\pi N_\uparrow}
  \\ \nonumber
  c_{S\uparrow} \left( x\right) &=& \frac{1}{\sqrt{2\pi\tau_c}}
  e^{i\Phi_{S\uparrow}\left(x\right)} e^{i\pi N_{S\uparrow}}
  \\  \nonumber
  c_{L\downarrow} \left( x\right) &=& \frac{1}{\sqrt{2\pi\tau_c}}
  e^{i\Phi_{L\downarrow}\left(x\right)} e^{i\pi N_{S\downarrow}}
  e^{i\pi N_\uparrow}
  \\  \nonumber
  c_{R\downarrow}\left(x\right) &=& \frac{1}{\sqrt{2\pi\tau_c}}
  e^{i\Phi_{R \downarrow}\left( x\right) } e^{i\pi \left(N_\uparrow
    +N_\downarrow \right)}
  \\ \nonumber
  c_{S\downarrow } \left( x\right) &=& \frac{1}{\sqrt{2\pi\tau_c}}
  e^{i\Phi_{S\downarrow} \left(x\right)} e^{i\pi N_{S\downarrow}}
  e^{i\pi N_\uparrow}
\end{eqnarray}
where the exponentials with the number operators
\begin{eqnarray}
  \label{Noperators}
  N_{\mu\sigma} &=& \int\! dx \,c^\dagger_{\mu\sigma}\!\left( x \right)
  c_{\mu\sigma}\! \left( x \right) \nonumber \\
 N_\sigma &=& \sum_\mu N_{\mu\sigma}
\end{eqnarray}
 have been inserted to insure anticommutation relations between the
various different fermion operators. Also, $\tau_c^{-1}$ is an cutoff energy
scale of the order the renormalized bandwidth.  In the above notation
the (normal ordered) local electron operators can be expressed simply
in terms of the bosonic variables
\begin{equation}
  \label{dPhidxdef}
  c_{\mu\sigma}^\dagger \!\left( x \right) c_{\mu\sigma} \!\left(
  x \right) = \frac{1}{2\pi} \frac{\partial\Phi_{\mu\sigma}
    \!\left( x \right)}{ \partial x}
\end{equation}
The electronic kinetic energy  can
also be represented in a standard manner in terms of bosonic
variables  $\phi_{\mu\sigma}\left( \epsilon \right)$ which obey the
canonical commutation relations
\begin{equation}
  \label{phicomrelation}
  \left[ \phi_{\mu\sigma}\left(\epsilon\right),
    \phi^\dagger_{\mu'\sigma'}\left(\epsilon'\right) \right] = 2\pi
    \delta_{\mu\mu'}
    \delta_{\sigma\sigma'} \delta\left( \epsilon-\epsilon'\right)
\end{equation}
with
\begin{equation}
  \label{Phimuasfunctionofphi}
  \Phi_{\mu\sigma} \left(x \right)= \int_0^{\infty}
  \frac{d\epsilon}{\sqrt{2\pi\epsilon}} \left[ \phi_{\mu\sigma}
  \left(\epsilon\right) e^{i\epsilon x} +\phi^\dagger_{\mu\sigma}
    \left(\epsilon\right) e^{-i\epsilon x}\right] e^{-
  \frac{\epsilon\tau_c}{2}},
\end{equation}
by
\begin{equation}
  \label{Ho(phi)}
  {\cal H}_0 = \sum_{\mu\sigma} \int_0^\infty \frac{d\epsilon}{2\pi}
  \,\epsilon\, \phi^\dagger_{\mu\sigma} \!\left( \epsilon \right)
  \phi_\mu \sigma \!\left( \epsilon \right) e^{-\epsilon\tau_c}
\end{equation}
with both integrals involving only positive energies (corresponding to
the fact that the electrons $c^\dagger_{\mu\sigma} \left( \epsilon
\right)$ are only left moving) and with a cutoff energy of the order
of the renormalized bandwidth $\tau_c^{-1}$. [Note that the energy
$\delta$-function in Eq(\ref{phicomrelation}) is rounded at energies
$\epsilon \approx \tau_c^{-1}$.] It is useful to define even and odd
combinations of the $L,R$ Bose fields
\begin{equation}
  \label{defPhieo}
  \Phi_{e,o \sigma}= \frac{1}{\sqrt{2}}\left(\Phi_{L\sigma} \pm
   \Phi_{R\sigma}\right).
\end{equation}
Then the most important parts of the effective Hamiltonian can be
written as
\begin{eqnarray}
  \label{defHeff}
  {\cal H}_{eff} &=& {\cal H}_0 + \frac{Q_1}{\sqrt{2}} \sigma_z \sum_\sigma
  \frac{\partial\Phi_{o\sigma}}{\partial x}
   \\ \nonumber
  &+& \frac{\Delta_1}{\tau_c}
  \sum_\sigma \left(\sigma_+
   \eta_\sigma \exp\left[i\sqrt{2}\Phi_{o\sigma}\right] +
   h.c.\right)
\end{eqnarray}
where the boson fields in the last two terms are evaluated at the
origin ($x=0$) and we consider later  the term proportional to
$\Delta_3$, which will turn out to be irrelevant. Here we
have defined the  factors
\begin{equation}
  \label{etapm}
  \eta_\sigma = e^{i\pi\left(N_{L\sigma}+ N_{R\sigma} \right)},
\end{equation}
which can take values $\pm 1$, since the number operators
$N_{\mu\sigma}$ can only take integer values. In the case that
$\eta_\sigma$ has a definite sign, (i.e. when $N_{L\uparrow}+
N_{R\uparrow}$ and $N_{L\downarrow}+ N_{R\downarrow}$ are separately
conserved) these factors can be neglected, as we shall do for the
moment. Note that at this point the $S$-electrons are decoupled; we
will later reintroduce terms that couple them to the $L$ and $R$
degrees of freedom.

Eq(\ref{defHeff}) is essentially equivalent to the Hamiltonian of the
anisotropic 2CK model, which has an $O(2)$ symmetry [$\Phi_{o\sigma}
\rightarrow \Phi_{o\sigma} + \theta/\sqrt{2}$, $\; \; \sigma_\pm
\rightarrow \sigma_\pm e^{\mp i\theta}$; $\; \; \Phi_{o\sigma}
\rightarrow -\Phi_{o\sigma}$, $\sigma_\pm \rightarrow \sigma_\mp$, $\;
\; \sigma_z \rightarrow - \sigma_z$]; this is sufficient to make it
flow to the $SU(2)$ symmetric 2CK fixed point.\cite{EK,Georges,MF2}

After decomposing the bosonic fields and $\eta_\sigma$ into charge and
spin components,
\begin{equation}
  \label{Phiecs}
  \Phi_{e\sigma}=\frac{1}{\sqrt{2}} \left(\Phi_{ec} + \sigma
  \Phi_{es}\right) \; \; \; \; \; \; \; \; \; \; \; \; \mbox{ $etc.$}
\end{equation}
and
\begin{equation}
  \label{etaeces}
  \eta_\sigma = e^{i\pi \left(N_{ec} + \sigma N_{es} \right)},
\end{equation}
with $\sigma=\pm$ for spin $\uparrow, \downarrow$, respectively, we
follow Emery and Kivelson,\cite{EK} and perform a unitary
transformation
\begin{equation}
  \label{Utrans1/2}
  U= \exp\left[ - \frac{i}{2} \sigma_z
  \Phi_{oc} \right]
\end{equation}
which transforms ${\cal H}_{eff}$ to the
form\cite{EK,MF2,Georges}
\begin{equation}
  \label{Heffbosonizedrotated}
  {\cal H'}_{eff} = U {\cal H}_{eff} U^+ = {\cal H}_0 + \left( Q_1 -
  \frac{1}{2} \right) \sigma_z \frac{\partial \Phi_{oc}
      }{\partial x}
  + \frac{2\Delta_1}{\tau_c}  \sigma_x  \cos\Phi_{os}.
\end{equation}
As was first pointed out by Emery and Kivelson\cite{EK} the above
Hamiltonian can be diagonalized exactly when $Q_1 = 1/2$ via
refermionizing the bosonic and pseudo-spin operators. This solution is
analogous to the Toulouse limit\cite{Toulouse} of the one-channel
Kondo model; here it captures the behavior of the 2CK model.  Sengupta
and Georges\cite{Georges} showed that $\sigma_z \frac{\partial
  \Phi_{oc} }{\partial x}$ is one of the leading irrelevant operators
with dimension 3/2, while a second dimension 3/2 operator was found by
us\cite{MF2} in the context of the 2CK model, consistent with
expectations from conformal field
theory.\cite{Affleck1,Affleck3,Affleck1a} The RG eigenvalues here are
one minus the dimension of the local operators.  There are no relevant
operators consistent with $O(2)$ [and spin $SU(2)$] symmetry.
However, the underlying symmetry of our system is $C_{3v}$, which is
lower than $O(2)$.  Although no extra relevant operators exist if we
lower the symmetry to $C_{3v}$, there will turn out to be one extra
irrelevant, dimension 3/2 operator allowed and a fourth, if electrons
$c^\dagger_A$ from the antisymmetric representation $\Gamma_2$ are
coupled to the impurity.

Unfortunately, the refermionization
technique does not work for all important operators allowed by
$C_{3v}$. Thus, in order to show that the above are indeed the only
allowed leading irrelevant operators, we analyze the problem with a
method recently used by Fabrizio and Gogolin\cite{Gogolin1} in the
context of the four-channel Kondo model following ideas of Fisher and
Zwerger\cite{MPA} and Schmid.\cite{Schmid}

\subsection{Analysis of effective Hamiltonian}

The approach described below is based on the observation that
$\Delta_1$ in Eq(\ref{Heffbosonizedrotated}) formally has dimension
1/2 and therefore will flow under RG to arbitrarily large values.  In
this case it makes sense to ``diagonalize'' this term first and treat
${\cal H}_0$ as a perturbation. Without loss of generality we assume
that $\Delta_1 > 0$. Neglecting ${\cal H}_0$, the minimum energy
values of $\Phi_{os} \left( x=0 \right)$ are, in the impurity basis
where $\sigma_x$ is diagonal, $\Phi_{os} \left( x=0 \right) = \left(
2n +1 \right) \pi$ when $\sigma_x= +1$ and $\Phi_{os} \left( x=0
\right) = 2n \pi$ when $\sigma_x= -1$. The presence of ${\cal H}_0$
has two effects. First, it includes (via Eqs
\ref{Phimuasfunctionofphi} and \ref{Ho(phi)}) the conjugate momentum
of $\Phi_{os} \left(0 \right)$, $\Pi_{os} \left( 0\right) = -
\frac{1}{2\pi} \frac{\partial \Phi_{os} \left( 0\right)}{\partial x}$.
Thus, in the semi-classical approximation, which is valid for large
$\Delta_1$, the phase $\Phi_{os} \left( 0 \right)$ will tunnel
occasionally between its various minimum energy values, with an
effective tunnelling rate proportional to $e^{-S_{kink}}$, where
$S_{kink}$ is the classical action of the motion of $\Phi_{os} \left(
0 \right)$ in the inverted potential\cite{SColeman} ($\Delta_1
\rightarrow - \Delta_1$), from, say $-\pi$ to $\pi$. The value of
$S_{kink}$, which is large when $\Delta_1$ is large, is not important
and clearly non-universal, since it depends on the details near the
cutoff energy.\cite{Schmid}

The other effect of ${\cal H}_0$ is via the coupling of $\Phi_{os}
\left( x=0 \right)$ to bosons located away from the impurity
  [$\Phi_{os} \left( x \neq 0\right)$], which will tend to {\em
    screen} the ``kinks'' (tunnelling events) of $\Phi_{os}\left( 0
\right)$. The underlying reason is that the ground states of
$\Phi_{os} \left( x \neq 0\right)$ with different values of
$\Phi_{os}\left( 0 \right)$ are mutually {\em orthogonal}, the
manifestation of Anderson's electron orthogonality catastrophe. To
analyze this effect, we integrate out the bosons $\Phi_{os} \left( x
\neq 0\right)$, leading to the following effective imaginary time
action for $\Phi_{os}\left( x=0, \tau \right) \equiv \Phi_{os}\left(
\tau \right)$
\begin{equation}
  \label{Seff}
  S_{eff}= \frac{1}{\left( 2\pi \right)^2} \int d \! \tau \int d \!
  \tau' \; \; \frac{\left( \Phi_{os} \left( \tau \right) - \Phi_{os}
  \left(  \tau' \right) \right)^2}{\left( \tau -\tau' \right)^2}
\end{equation}
with $\Phi_{os}\left( \tau \right)$ typically being a series of widely
separated (on the cutoff scale) kinks between different minima. This
action has been studied extensively in a variety of quantum impurity
problem contexts\cite{MF1,Gogolin1,MPA,Schmid,Gogolin2} and it is
straightforward to show that the operators ${\cal O}_m$, corresponding
to $\Phi_{os}\left( \tau \right)$ hopping by $m\pi$, have dimension
$\frac{m^2}{2}$. Therefore, when $\sigma_x$ has no dynamics, the
leading irrelevant operator is ${\cal O}_2$, corresponding to hops
between adjacent minima of the $\cos\Phi_{os}$ term.

\subsection{Leading and other Irrelevant Operators}

In the absence of relevant operators, at temperatures small compared
to $T_K$ the system will flow towards the fixed point. The flow close
to the fixed point will be dominated by the leading irrelevant
operators, i.e. the ones with the least negative RG eigenvalue, since
other operators, being more irrelevant, will renormalize to zero
faster and therefore will become negligible. At non-zero temperatures
the RG flows will be terminated when the renormalized bandwidth
becomes of order the temperature. At this scale the only
non-negligible processes that will enter in physical quantities are
the leading irrelevant ones and the temperature dependence of their
renormalized amplitudes will determine the low temperature behavior of
the system.

First we analyze the leading irrelevant operators with dimension 3/2.
As mentioned above, the first is $\sigma_z \partial\Phi_{oc}/\partial
x$. Since $\Phi_{oc}$ is decoupled from $\Phi_{os}$ and $\sigma_x$
when $Q_1= 1/2$, the derivative term clearly has dimension 1. The
effect of $\sigma_z$ (or, equivalently $\sigma_y$) is non-trivial:
since it anticommutes with $\sigma_x$ in the $\Delta_1$ term, it
effectively reverses the sign of $\sigma_x$.\cite{Gogolin1} As a
result, the ground state value of $\Phi_{os}$ shifts by $\pi$ when
$\sigma_z$ is applied, thereby giving $\sigma_z$ (and $\sigma_y$)
dimension 1/2, since it forces a $\pi$ kink in $\Phi_{os}$, just like
${\cal O}_1$ above. As a result, the operator proportional to ($Q_1-
1/2$) has overall dimension 3/2, in agreement with other
approaches.\cite{Georges,MF2}

As discussed in Ref. \cite{MF2}, there is a second dimension 3/2
operator in the spin sector, which in the original electron operator
representation is
\begin{eqnarray}
  \label{2ndleadirropfermionrep}
  d_L^\dagger d_R \left[ c_{R\uparrow}^\dagger c_{L\downarrow} \left\{
  c_{L\downarrow}^\dagger c_{L\uparrow} + c_{R\downarrow}^\dagger
  c_{R\uparrow} \right\} + c_{R\downarrow}^\dagger
  c_{L\uparrow} \left\{ c_{L\uparrow}^\dagger c_{L\downarrow} +
  c_{R\uparrow}^\dagger c_{R\downarrow} \right\} \right. \\ \nonumber
  \left. \left\{ c_{L\uparrow}^\dagger  c_{L\uparrow} +
  c_{R\uparrow}^\dagger c_{R\uparrow} - c_{L\downarrow}^\dagger
  c_{L\downarrow} -
  c_{R\downarrow}^\dagger c_{R\downarrow} \right\} \left\{
  c_{R\uparrow}^\dagger c_{L\uparrow} - c_{R\downarrow}^\dagger
  c_{L\downarrow} \right\} \right] +h.c.,
\end{eqnarray}
written in an explicitly spin-$SU(2)$ invariant form. Note that this
operator involves two electron processes [part of ${\cal
  H}_{hop}^{(2)}$] and thus is irrelevant for
weak coupling. After bosonizing and performing the unitary
transformation of
Eq(\ref{Utrans1/2}), the most relevant part in the square brackets
becomes
\begin{equation}
  \label{2ndleadingirrop}
  \sigma_y \sin\Phi_{os} \frac{\partial \Phi_{es}}{\partial x}.
\end{equation}

Apart from the above operators there is an extra leading irrelevant
operator consistent with the $C_{3v}$ symmetry. Similar to the above,
in the original fermion representation, this operator has the
spin-$SU(2)$ invariant form
\begin{equation}
  \label{irrelop1}
  d_L^\dagger d_R \left[ c_{S\uparrow}^\dagger c_{L\uparrow}
  c_{R\downarrow}^\dagger c_{S\downarrow} + c_{R\uparrow}^\dagger
  c_{S\uparrow}
  c_{S\downarrow}^\dagger c_{L\downarrow} - \frac{1}{2} \left(
  c_{R\uparrow}^\dagger c_{L\uparrow} - c_{R\downarrow}^\dagger
  c_{L\downarrow} \right) \left( c_{S\uparrow}^\dagger c_{S\uparrow} -
  c_{S\downarrow}^\dagger c_{S\downarrow} \right) \right] + h.c. \; .
\end{equation}
After bosonizing and performing the unitary transformation of
Eq(\ref{Utrans1/2}), the last term in the square brackets becomes
proportional to
\begin{equation}
  \label{irrelop1'}
  \sigma_y \sin \Phi_{os} \frac{\partial \Phi_{Ss}}{\partial x}.
\end{equation}
exactly analogous to Eq(\ref{2ndleadingirrop}).  Since $\sigma_y$
flips the sign of $\sigma_x$ in the $\Delta_1$-term and $\Phi_{os}$ is
essentially frozen, Eqs(\ref{irrelop1'}) and (\ref{2ndleadingirrop})
have dimension 3/2. Note that although nominally $\sin\Phi_{os} = 0$
{\em at} the minima of the cosine potential, the kink forced by the
$\sigma_y$ gives $\sin\Phi_{os}$ a non-zero value at the kink.

It is important to comment that the extra dimension 3/2 operator
(Eq(\ref{irrelop1'})) only exists because of the presence of the
symmetric electron state $c^\dagger_{S\sigma}$, which is invariant
under $C_{3v}$. In fact, {\em any} $C_{3v}$ singlet fermion bilinear
made up of electrons that can couple to the impurity through its spin,
like $c_S$ here, will produce a dimension 3/2 operator. In particular
the operator similar to Eq(\ref{irrelop1}), but with the
$c_S$-electrons replaced by the $c_A$ electrons of the odd
one-dimensional $ \Gamma_2 $-representation, will also have dimension
3/2. Note however, that due to their symmetry, $\Gamma_2$-electrons
tend to interact more weakly with the impurity.

The first two terms in Eq(\ref{irrelop1}),
transformed via Eq(\ref{Utrans1/2}), become proportional to
\begin{equation}
  \label{irrelop2'}
  \sigma_x \cos\left[ \Phi_{es} - \sqrt{2} \Phi_{Ss} \right]
\end{equation}
where we have neglected the exponential factors $\exp\left[ i \left(
N_{ec} \pm N_{es} \right) \right]$. Naively this operator also has
dimension 3/2. However, due to the presence of $\Phi_{es}$, we need to
be more careful with the $\exp\left[ i \left( N_{ec} \pm N_{es}
\right) \right]$ in the $\Delta_1$ term of ${\cal H}'_{eff}$,
Eq(\ref{Heffbosonizedrotated}).  Including these factors explicitly
alters the dominant terms in ${\cal H'}_{eff}$ to
\begin{equation}
  \label{Delta1'}
 {\tilde{\cal H}}_{eff} = {\cal H}_0 +  \frac{2\Delta_1}{\tau_c}
  \left( \sigma_x \cos \pi N_{ec} + \sigma_y
  \sin \pi N_{ec} \right) \cos\left[ \Phi_{os} -\pi N_{es} \right]
\end{equation}
Thus the presence of an $\exp\pm i\Phi_{es}$ factor in
Eq(\ref{irrelop2'}) causes a sign change of the $\Delta_1$ term since
it takes $N_{es} \rightarrow N_{es} +1$ and therefore behaves just as
$\sigma_z$ does, increasing the dimension of the operator to
2. Similar arguments hold for subleading terms from
Eq(\ref{2ndleadirropfermionrep}).

Naively one would expect all parts of operators such as
Eqs(\ref{2ndleadirropfermionrep}) and (\ref{irrelop1}) to contribute
to the {\em leading} irrelevant operators since the they are orbital
and spin singlets. The reason they do not is that through the
bosonization scheme implemented above, in addition to the unitary
transformation of Eq(\ref{Utrans1/2}), the underlying orbital $SU(2)$
symmetry of the Hamiltonian is broken down to an $O(2)$
symmetry\cite{MF2} (which is the symmetry of the XXZ anisotropic
orbital Kondo model\cite{EK}). Therefore some parts of the various
operators will mix will orbital-spin-2 operators associated with this
symmetry breaking, thereby acquiring different scaling dimensions.
Since the orbital-spin-2 operators have dimension 2 at the orbitally
$SU(2)$ invariant fixed point and are thus more irrelevant than the
dimension 3/2 leading irrelevant operators, the breakdown of the
orbital $SU(2)$ symmetry to an $O(2)$ symmetry will not affect the low
temperature behavior, as the dimension 2 operators will become
negligibly small at low temperatures. This is the underlying reason
why the solvable point of the 2CK model found by Emery and
Kivelson\cite{EK} captures correctly the singular low temperature
behavior of the 2CK fixed point.\cite{Georges,Ye}

Other than $\Delta_1$ and $Q_1$, the only one-electron impurity
operator that survives when $d_S$ is suppressed, is $\Delta_3$.  The
most relevant part of this near the 2CK fixed point can be written as
\begin{equation}
  \label{irrelop3'}
  \sigma_x \cos\left[ \frac{\Phi_{Ss}}{\sqrt{2}} - \frac{\Phi_{es} +
  \Phi_{os}}{ 2} \right] \cos \left[ \frac{\Phi_{Sc}}{\sqrt{2}} -
  \frac{\Phi_{ec} + 3 \Phi_{oc}}{ 2} \right].
\end{equation}
Naively this has dimension 15/8, since $\Phi_{os}$ is frozen. However,
the presence of $\exp\pm i\frac{ \Phi_{es}}{2}$ shifts $N_{es}$ in
Eq(\ref{Delta1'}) by $\pm 1/2$, turning the cosine into a sine. This
forces a shift of $\Phi_{os}$ by $\pm \pi/2$, thereby increasing the
dimension of the $\Delta_3$ operator in Eq(\ref{irrelop3'}) by 1/8 to
2. This is in complete agreement with the dimension of $\Delta_3$
derived using conformal field theory methods. To see this, we observe
that this operator combines $c_{S\sigma}$ or $c^\dagger_{S\sigma}$,
which are free Fermi fields, uncoupled from the impurity at the fixed
point, and hence having dimension 1/2, to $\left( d^\dagger_L d_R
c^\dagger_{L\sigma} + d^\dagger_R d_L c^\dagger_{R\sigma} \right)$ and
$\left( d^\dagger_L d_R c_{R\sigma} + d^\dagger_R d_L c_{L\sigma}
\right)$, respectively. These operators by themselves would break the
$U(1)$-charge symmetry of the model; since they involve an odd number
of electron operators, they have charge 1. [In our bosonization
scheme, this is manifested in them changing sign under $\Phi_{ec}
\rightarrow \Phi_{ec} + 2 \pi$]. Furthermore, they clearly transform
as a spin-doublet. The dimension of a charge-1, spin-1/2 operator at
the 2CK fixed point is $n$ + 1/2, with $n$ a non-negative
integer.\cite{JvDthesis}
However the operators with $n=0$ do not respect the $C_{3v}$ symmetry,
in particular, unlike the above operators, they are not invariant
under $c_L \rightarrow c_L \exp \left[ \frac{2\pi i}{3} \right]$ and
$c_R \rightarrow c_R \exp \left[- \frac{2\pi i}{3} \right]$, since
they transform as doublets under $C_{3v}$. As a result, the allowed
operators have $n=1$ so that the scaling
dimension of $\left( d^\dagger_L d_R c^\dagger_{L\sigma} + d^\dagger_R
d_L c^\dagger_{R\sigma} \right)$ and $\left( d^\dagger_L d_R
c_{R\sigma} + d^\dagger_R d_L c_{L\sigma} \right)$ is 3/2; combined
with the 1/2 of $c_{S\sigma}$ or $c^\dagger_{S\sigma}$, this makes the
dimension of $\Delta_3$ equal to 2, as found above.

By similar arguments, it can be shown that the dimensions of all
physical operators are integers or half integers.  This should be
expected from the conformal field theory point of view, since the 2CK
fixed point only has boundary operators with half-integer
dimension.\cite{JvDthesis} Combining these with free fermions, such
as $c_{S\sigma}$, that couple only through their spin (which plays the
role of the ``flavor'' in the spin 2CK model), does not alter this result, as
we saw explicitly in the case of $\Delta_3$.

\subsection{Physical consequences}

The low temperature behavior of the impurity and the electrons that
couple to it will be governed by the 2CK fixed point discussed above
and the flow towards the fixed point.  In the absence of the
irrelevant operators, the effective Hamiltonian of
Eq(\ref{Heffbosonizedrotated}) with $Q_1 = 1/2$, i.e.
Eq(\ref{Delta1'}), couples only the $\Phi_{os}$ Bose fields to the
impurity. [Although strictly speaking this is not the fixed point
Hamiltonian, as discussed in Ref. \cite{MF2} it has the same essential
structure, differing from the fixed point primarily in $\Delta_1$
being finite.]

This means that quantities like the spin susceptibility, $\chi$, which
involve only other linear combinations of the Bose fields--in this
case correlations of the spin operators from electrons in different
representations, e.g.  $\frac{\partial \Phi_{es} \left( 0 \right)
  }{\partial x} $ --will decouple from the impurity and be
non-singular. However the leading irrelevant operator in the spin
sector Eq(\ref{2ndleadingirrop}) can combine with $\frac{\partial
  \Phi_{es}}{\partial x}$ to yield $\sigma_y \sin\Phi_{os}$, which has
dimension 1/2 as discussed above.  Similarly, the other two leading
irrelevant operators involving $S$-electrons and $A$-electrons can
also yield $\sigma_y \sin\Phi_{os}$. As shown by Sengupta and
Georges,\cite{Georges} this effect gives rise to a contribution to the
susceptibility
\begin{equation}
\label{chi}
  \Delta \chi \sim \ln \left(T_K/T \right)
\end{equation}
at low temperatures, with the coefficient being proportional to the
sum of the squares of the coefficients of these three leading
irrelevant operators in the effective low energy Hamiltonian.

On the other hand, all dimension 3/2 operators, including the one
proportional to $(Q_1-1/2)$, will contribute to the specific heat
$C_V$, giving a singular correction\cite{Georges}
\begin{equation}
  \label{CV}
  \Delta C_V \sim T \ln \left(T_K/T \right)
\end{equation}
with the coefficient being proportional to the sum of squares of all
four leading irrelevant operators. As a result, the normalized ratio
of $T \Delta \chi$ to $\Delta C_V$--the Wilson ratio--will not be
universal. However, due to the similarity in their form, the three
operators entering  the susceptibility, $\frac{ \partial
  \Phi}{\partial x} \sigma_y \sin \Phi_{os}$, with the $\Phi$ in the
derivative being one of $\Phi_{es}$, $\Phi_{Ss}$ and $\Phi_{As}$, will
enter the specific heat in exactly the same combination.
Thus for $Q_1 =1/2$ the normalized ratio of the specific heat over the
susceptibility will be universal. As a consequence, to find a general
universal ratio, one extra independent measurable quantity is
required, that involves, e.g. only $\left( Q_1 -1/2\right)^2$.
It should be noted that the contribution to the resistivity which
scales as\cite{Affleck1,Affleck3}
\begin{equation}
  \label{rho}
  \Delta \rho \sim \sqrt{T}
\end{equation}
is not suitable, because it involves a sum over the amplitudes of all
the leading irrelevant operators.

\subsection{Symmetry breaking perturbations}

The most important symmetry breaking perturbation is a strain or other
anisotropy that favors one of the three impurity positions. This most
simply appears in the Hamiltonian as in Eq(\ref{C3vbreakingterms})
where the phase of $M$ determines the preferred direction for the
impurity relative to the center of symmetry. This corresponds to
$\sigma_x Re M + \sigma_y Im M$ in the effective Hamiltonian,
operators which have dimension 1/2 and are hence relevant, albeit less
so than from naive power counting. At energy scales above $T_K$, $M$
just grows linearly with energy [as described in Section
\ref{sec:IVc3v}.1], but on scales below $T_K$, it grows more slowly
with eigenvalue 1/2.  The perturbation thus becomes important and
destroys the non-Fermi-liquid behavior below the energy scale
\begin{equation}
  \label{TM}
  T_M \sim \frac{\left| M \right|^2}{T_K},
\end{equation}
the $\left| M \right|^2$ reflecting the eigenvalue of 1/2. Thus the
novel 2CK behavior is more stable than might be expected. Note that
similar crossovers are found in the presence of spin and electron-
orbitally-coupled magnetic fields since these break the spin symmetry
or the $L,R$ symmetry, respectively.

It should be noted that the operator $M$ in Eq(\ref{C3vbreakingterms})
can also be generated through electronic processes that break the
triangular symmetry caused, e.g., by the presence of static impurities
in the vicinity of the tunnelling impurity. These give rise to mixing
between the electronic wavefunctions interacting with the tunnelling
impurity or between these and other wavefunctions.\cite{Wingreen} This
mixing breaks the $C_{3v}$ symmetry. In the presence of
electron-assisted-tunnelling processes, such as $A_1, \ldots, A_6$
discussed before Eq(\ref{Vdiag}), these $C_{3v}$ symmetry breaking
electronic terms, such as
\begin{equation}
  \label{staticbreaking}
  \sum_\sigma \left(P c^\dagger_{L\sigma} c_{R\sigma} +
  P^* c^\dagger_{R\sigma} c_{L\sigma} \right)
\end{equation}
will generate under RG the operator $M$.

\subsection{Effects of higher symmetries}

Impurities with a $\Gamma_3$-doublet ground state tunnelling around a
point in the lattice with higher symmetry, such as octahedral symmetry
in a cubic lattice, will clearly not have any relevant operators at
the 2CK fixed point. The reason is that since these relevant operators
are not allowed by $C_{3v}$, a subgroup of the octahedral group $O$,
they will not be allowed by $O$ either. However, there will be extra
irrelevant operators, which turn out also to have dimension 3/2,
therefore not essentially changing the low temperature physics of the
2CK fixed point. Electrons from representations $\Gamma$ such that
$\Gamma \otimes \Gamma$ contains the trivial representation, will also
have an associated dimension 3/2 operator of the form of
Eq(\ref{irrelop2'}).  Furthermore, since $\Gamma_4 \otimes \Gamma_4$
and $\Gamma_5 \otimes \Gamma_5$ contain, apart from $\Gamma_1$, also
the doublet $\Gamma_3$, electrons in these representations can couple
orbitally with the impurity, in the form $d^\dagger_L d_R \sum_\sigma
c^\dagger_{\Gamma_i \sigma} c_{\Gamma_i \sigma}$, for $i=4, 5$, giving
two additional dimension 3/2 operators.  As a result, in the case of
octahedral symmetry $O$, where there are two 3-dimensional
representations of electrons, $\Gamma_4$ and $\Gamma_5$, there will be
{\em four} extra dimension 3/2 operators, leading to a total of eight
leading irrelevant operators.

If the impurity is in a {\em triplet} ground state ($\Gamma_4$ or
$\Gamma_5$ in the octahedral, or $T$ in the tetrahedral case), due to
arguments similar to those in this paper, it is anticipated that there
will be a stable non-Fermi-liquid fixed point, analogous to, but
different from, the 2CK fixed point for an impurity $\Gamma_3$-doublet
ground state. In this case the situation is similar to the critical
point of the doublet case at which ${\cal E}$ changes sign,
corresponding roughly to a triplet impurity ground state;\cite{MS}
this is briefly discussed in the next Section.

\section{Critical Manifold separating the 2CK from the Fermi liquid
  fixed points}
\label{sec:VIc3v}

\begin{figure}[htpb]
    \begin{center}
      \leavevmode
      \epsfxsize=4truein
      \epsfbox{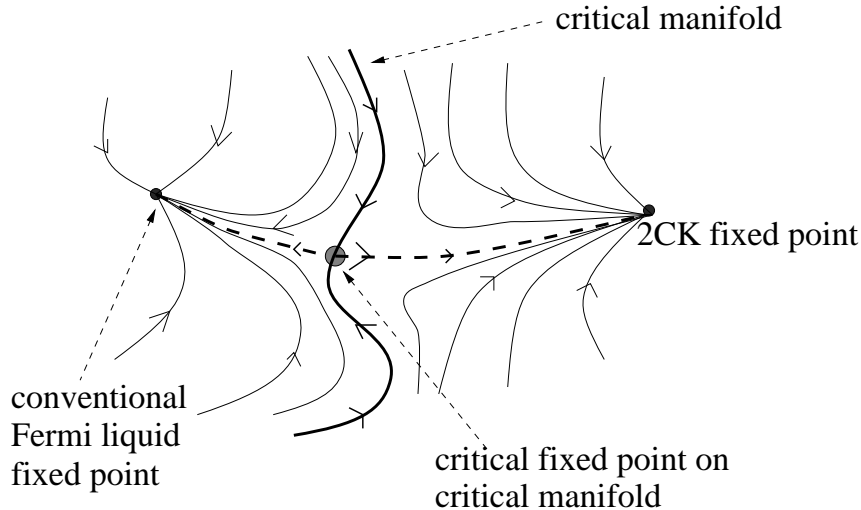}
    \end{center}
    \caption[Renormalization group flows]{Schematic
      of renormalization group flows for the three site problem at
      zero temperature. There are two stable fixed points, namely the
      non-Fermi-liquid 2CK fixed point and the conventional
      Fermi-liquid fixed point, at which the impurity ground state is
      a $\Gamma_1$-singlet which is completely screened at low
      temperatures. The critical manifold separating the basins of
      attraction of the two stable fixed points is also shown.  The
      flow on this manifold goes to a critical fixed point, which has
      higher orbital symmetry. The critical fixed point has one
      unstable direction that respects $C_{3v}$; the dashed lines show
      the universal crossover flows in this unstable direction from
      the unstable to the stable fixed points.}
     \label{fig:fixedpoints}
\end{figure}

For all of the physics discussed so far, the sign of ${\cal E}$ is
crucial. Here we consider what happens if ${\cal E}$ can be made to
change sign by a change in the system, for example by applying
pressure. From the analysis of Section \ref{sec:IVc3v}, it should be
clear that it is really the sign of the effective hopping,
${\tilde{\cal E}}$, which includes renormalizations due to the effects
of the electron assisted tunnelling, etc., that is important.
Generally there will be a critical manifold separating the novel 2CK
low temperature behavior from a conventional Fermi liquid regime in
which the impurity is in its non-degenerate symmetric state with
simply potential scattering of the electrons. This is schematically
shown in Fig.\ref{fig:fixedpoints}. Since both the fixed points are
stable at zero temperature, the critical manifold separating their
domains of attraction can generically be crossed by a one parameter
change in the system.

When the system lies on the critical manifold, which corresponds to
${\cal E} = 0$ at small coupling, it will presumably flow to a new
unstable fixed point. We conjecture that, apart from the charge-$U(1)$
and spin-$SU(2)$ symmetries, this has an additional orbital-$SU(3)$
symmetry, where the $L$, $R$, $S$ triplet is identified with the
fundamental representation of $SU(3)$. The behavior of the system on
the critical manifold and close to this fixed point will be analyzed
elsewhere,\cite{MS} but below we quote some of the important results.

Within the critical manifold this fixed point is stable and the RG
flows close to the fixed point are determined from the dimension of
the leading irrelevant operator, which in this case is 7/5. Therefore
the impurity corrections to the specific heat, resistivity and
susceptibility at low temperatures are $\Delta C_V \sim T^{4/5}$,
$\Delta \rho \sim T^{2/5}$ and $\Delta \chi \sim T^{-1/5}$,
respectively.  Note that although the leading irrelevant operator and
the symmetry of this fixed point are the same as those of the
three-channel Kondo model,\cite{Affleck4} the fixed point itself and
the operator content are not the same. This can easily be seen by
noting that while in the three-channel Kondo model an impurity
$SU(2)$-doublet couples to 3 channels of electrons through their spin,
in this case the impurity is an $SU(3)$-triplet coupling to 2 channels
of electrons through their orbital $SU(3)$ index.

Deviations away from the critical manifold are relevant dimension 3/5.
In a $C_{3v}$ symmetric situation only one such operator exists, which
can be represented by $(2 d^\dagger_S d_S - d^\dagger_L d_L -
d^\dagger_R d_R)/3$. This is therefore analogous to ${\cal E}$ and it
governs the RG flows away from the critical fixed point toward the
Fermi-liquid and the 2CK stable fixed points.  Therefore, tuning only
one parameter (such as external pressure) the novel non-Fermi-liquid
fixed point associated with this higher orbital symmetry might be
observed. In particular, close to the critical pressure, there will be
a low temperature regime with the critical behavior crossing over
below a temperature
\begin{equation}
  \label{TXpressure}
  T_X \sim \left| p- p_c
\right|^{5/2}
\end{equation}
to one of the two possible low temperature behaviors.

It should be noted that the above behavior also holds for the case of
spinless impurities with a {\em triplet} degenerate ground state in
tetrahedral or octahedral sites in a cubic crystal. In that case there
will be no unstable directions close to the fixed point in the absence
of symmetry breaking terms, just as for the 2CK fixed point in the
case of $C_{3v}$, and the low temperature physics would be governed by
the critical fixed point and its corresponding non-Fermi-liquid
behavior.

\section{Conclusions and Relevance to Experiments}
\label{sec:VIIc3v}

We finally discuss briefly the possible relevance of impurity
tunnelling to the nanoconstriction experiments of Ralph and
Buhrman,\cite{Ralph1,Ralph2,Buhrman} as well as potential future
experiments.

In Ref. \cite{MF2} we showed that the interpretation of the observed
non-Fermi-liquid-like behavior of these experiments as a defect
tunnelling between two sites\cite{Ralph2} runs into trouble due to the
existence of {\em two} relevant operators, that drive the system away
from the 2CK fixed point and therefore make it highly improbable that
a typical impurity could lie close enough to the critical manifold for
the non-Fermi-liquid behavior to be observable.  Another possible
interpretation is in terms of an interstitial, vacancy or other
impurity in the copper lattice structure which has a
$\Gamma_3$-degeneracy. In such a situation, the relevant perturbations
would only be induced by strains or other such effects; these tend to
split the doublet, as discussed above.

In the copper nanoconstriction experiments\cite{Ralph1,Ralph2} the
apparent Kondo temperatures are quite large, approximately $T_K\approx
10-15K$. This suggests that the impurity-electron interactions are
appreciable and thus the Kondo temperature, within our model, would be
the energy scale at which ${\cal E}$ becomes large. [As seen in
Section \ref{sec:IVc3v}, the RG eigenvalue of ${\cal E}$ should be
less than one. However, for not too large $V_1$ and $V_2$ a lower
bound for ${\cal E}$ is $T_K$.]  Such a fast tunnelling rate suggests
that a hopping process might involve either the slight rearrangement
of several atoms, as this can have a lower effective mass, or a small
displacement of an impurity (or impurity complex) from its central
symmetric position.

In the copper nanoconstriction devices\cite{Ralph2} no significant
deviation from the $\sqrt{T}$ temperature dependence of the impurity
conductance signal has been seen down to $50mK$, the lowest
temperature studied.  From Eq(\ref{TM}) one can therefore estimate
that the splitting energy scale $|M|$ would have to be less than
$0.7K$, i.e. less than $0.05 {\cal E}$. In more recent experiments in
$Pd$ and $V$\cite{Buhrman} a larger crossover temperature of about
$1.5K$ was observed, which would correspond to $M \sim 5K $, which is
about $0.3 T_K$. The stronger splitting in these materials might be
attributed to the fact that the nanoconstrictions are more stressed
than the corresponding copper devices\cite{Buhrman}.

In order to see whether a scenario involving impurities with
triangular symmetry might be reasonable, we need a rough estimate of
the magnitude of the effects of strain.  To be concrete, we assume a
Jahn-Teller potential, which in polar coordinates about the center of
the triangle has the form
\begin{eqnarray}
  \label{VJT}
  {\cal V}_{JT} (r) &=& \frac{A}{2} r^2 - B r^3 \cos 3 \theta + C r^4
  \\ \nonumber
  &-& \Delta r - \Delta' r^2 \cos 3 \theta
  - e \cdot D \cdot {\bf r}
\end{eqnarray}
with $e$ the strain tensor and $\Delta$ and $\Delta'$ respectively the
linear and quadratic splittings due to the Jahn-Teller effect. [Note
that the potential for a Jahn-Teller impurity with cubic symmetry that
has a doubly degenerate ground state would have a similar form to
Eq(\ref{VJT}). In this case ${\bf r}$ would be a tensor quantity, {\em
  even} under inversion, that parameterizes the deviations of atoms
close to the impurity from their equilibrium positions.] If $r$ is
measured in units of the lattice constant, $a$, then $A$, $B$, $C$ and
the third rank tensor $D$ are all of order elastic energies $10^5K$.
In the absence of strain, there will be three minimum energy positions
of the impurity with, for $\Delta$ and $\Delta' \sim \Delta/a$ small,
$r \approx r_0 \approx \Delta/A$. These will be separated by barriers
of order $\left( r_0/a \right)^3 10^5K$, with comparable contributions
arising from both the second and fifth terms in Eq(\ref{VJT}). The
tunnelling rate will be
\begin{equation}
  \label{EasfnofS}
  {\cal E} \sim \omega_D \sqrt{\frac{r_0}{a}} e^{-S}
\end{equation}
with $\omega_D$ the Debye frequency and $S$ the tunnelling action.
Thus for ${\cal E}$ to be about $10K$, the action $S$ cannot be very
large.

The dominant effects of strain arise in two ways: first, the
tunnelling rates between the three sites can become asymmetric. Since
some of the barriers will be changed by of order $eDr_0$ from
Eq(\ref{VJT}), we see that the effective $M$ resulting from this will
be
\begin{equation}
  \label{Mtun}
  M_{tunnelling} \sim \delta {\cal E} \sim {\cal E}
  \frac{eDr_0}{Br_0^3} \sim {\cal E} e \left( \frac{a}{r_0} \right)^2
\end{equation}
which decreases with $r_0$. In contrast, the direct splitting from the
change in energy of the minima is
\begin{equation}
  \label{Msplit}
  M_{split} \sim eDr_0 \sim e \left( \frac{r_0}{a} \right) 10^5 K,
\end{equation}
which grows with $r_0$.  For both of these to be less than $1K$, one
would need $r_0 \approx 0.1$\AA\ and $e\sim 10^{-4}$. So small a
strain appears implausible in a nanoconstriction. Thus, unless an
impurity configuration with anomalously weak coupling to strain or
with other special properties were to exist, this explanation for the
experiments of Ralph {\em et al}\cite{Ralph2,Buhrman} seems unlikely.

Other possible interpretations exist, in particular, a possible
explanation of the data was proposed by Wingreen {\em et
  al}\cite{Wingreen} that involves the existence of static randomness
in the neighborhood of the constriction. But overall, the
interpretation of these experiments is controversial at this time.

But impurity tunnelling about symmetric positions in metals might
occur in other situations. For example, it has been known for some
time\cite{Schilling,Dederichs} that undersized impurities often form
rather stable complexes with interstitials. These complexes, in $fcc$
lattices, form octahedral configurations. Thermally activated hopping
between equivalent configurations of these complexes has been
experimentally demonstrated at low temperatures in the case of dilute
$Co_x Al_{1-x}$ alloys,\cite{Vogl} which suggests that low barriers
might exist, even though in this system the tunnelling rates are
probably very low.\cite{Vogl}

In conclusion, we have shown that an impurity in a metal tunnelling
between positions with high symmetry--in particular, triangular
symmetry--can lead to generic non-Fermi-liquid behavior, which can be
characterized by the two-channel Kondo model. Symmetry breaking
perturbations were shown to be partially screened, thereby reducing
their effects and widening the potential temperature range for
observation of the novel 2CK behavior. In addition, we have shown that
by tuning a single parameter, such as pressure, the impurity behavior
can change to that of a conventional Fermi liquid, but at the critical
pressure the behavior would be controlled by a fixed point that is
conjectured to have a higher $SU(3)$ orbital symmetry.
A detailed analysis of the behavior on and close to this critical
manifold will be discussed in a future publication.\cite{MS}

\acknowledgments
We wish to thank Anirvan Sengupta, Bert Halperin, Andreas Ludwig, Paul
Fendley, Jan von Delft and Sharad Ramanathan for useful discussions
and Frans Spaepen for bringing references \onlinecite{Schilling} and
\onlinecite{Dederichs} to our attention.  This work was supported by the
National Science Foundation via Grants No. DMR 91-06237 and DMR
96-30064.


\end{document}